%% file: msurvey0207.tex
\documentclass[11pt]{amsart}
\usepackage{latexsym,amssymb,amsmath}
\usepackage{amsmath,amsthm,amssymb,amscd}
\usepackage[mathscr]{eucal}
\usepackage{verbatim}
\usepackage{epsfig}
\usepackage{color}
\usepackage{epsf,float}
\input cortdefs.tex

\swapnumbers  %% this puts sect number before "Theorem" etc.

\def\aa#1#2{a^{#1}_{#2}}
\def\bb#1#2{b^{#1}_{#2}}
\def\cc#1#2{c^{#1}_{#2}}
\def\tinf{{\rm inf}}
\def\subsmooth{{}_{smooth}}
\def\tbrank{{\underline{\bold R}}}
\def\trank{{\rm Rank}}\def\tspan{{\rm Span}}
\def\trankc{{ \bold R}}
\def\tlker{{\rm Lker}}\def\tliminf{\underline{\rm lim}}
\def\trker{{\rm Rker}}
\def\tlength{{\rm length}}

\def\uV{{\underline V}}
\def\aaa{{\bold {a}}}\def\bbb{{\bold {b}}}\def\ccc{{\bold {c}}} 
\def\tbase{{\rm Zeros}}

\begin{document}

\title{Geometry and the complexity of matrix multiplication}
\author{J.M. Landsberg}
%\date{July 13, 2004}
\begin{abstract}We survey   results in algebraic
complexity theory, focusing
on matrix multiplication. Our goals are  
 (i.) to show how   open questions  in algebraic
complexity theory are naturally
posed as questions in geometry and representation theory, 
(ii.) to motivate researchers   to work on these questions, and 
(iii.) to point
out relations with  more general problems
in geometry. The key geometric
objects for our study
are the secant varieties of Segre varieties.
We explain how these varieties are also
useful for algebraic statistics,   the study of
phylogenetic invariants, and quantum computing.
\end{abstract}
%\footnote{Supported by NSF grant DMS-0305829}
%\footnote{MSC 68Q17}
\email{jml@math.tamu.edu}

\maketitle
\section{Introduction}\label{section1}
\subsection{Strassen's algorithm}
\label{intro} 
Let $A$ and $B$ be $2\times 2$ matrices
$$
A=\begin{pmatrix}a^1_1 & a^1_2 \\ a^2_1
&a^2_2\end{pmatrix},\ \
B=\begin{pmatrix}b^1_1 & b^1_2 \\ b^2_1
&b^2_2\end{pmatrix}.
%,\ \ 
%C=\begin{pmatrix}c^1_1 & c^1_2 \\ c^2_1
%&c^2_2\end{pmatrix}
$$

Recall the usual algorithm to calculate
the matrix product $C=AB$:
\begin{equation}
\begin{aligned}\label{stdalg}
\cc 11&= \aa 11\bb 11+\aa 12\bb 21,\\
\cc 12&= \aa 11\bb 12+\aa 12\bb 22,\\
\cc 21&= \aa 21\bb 11+\aa 22\bb 21,\\
\cc 22&= \aa 21\bb 12+\aa 22\bb 22.
\end{aligned}
\end{equation}
This algorithm uses $8$ multiplications and
for  $n\times n$ matrices it uses $n^3$.

\donote{Question:} Is there a \lq\lq better\rq\rq\  algorithm for multiplying
matrices? By \lq\lq better\rq\rq\  one could mean an
algorithm that uses fewer
arithmetic operations ($+,-,*$), or simply fewer multiplications.
The number of multiplications needed governs the total
number of arithmetic operations in such a way
that asymptotic results   depend primarily
on the number of multiplications used. 
(See  Definition \ref{mexponentdef}   for 
a precise statement.) In this article we   focus
  exclusively on minimizing multiplications. 
(In actual implementations  memory cost
is also an important factor.)

In 1969 Strassen \cite{Strassen493} made the following   discovery.
Set
\begin{align*}
I&= (\aa 11 + \aa 22)(\bb 11 + \bb 22),\\ 
II&=(\aa 21 + \aa 22)\bb 11, \\
III&= \aa 11(\bb12-\bb 22)\\
IV&=\aa 22(-\bb 11+\bb 21)\\
V&=(\aa 11+\aa 12)\bb 22\\
VI&= (-\aa 11+\aa 21)(\bb11 +\bb12),\\
VII&=(\aa 12 -\aa 22)(\bb 21 + \bb 22),
\end{align*}

Now check for yourself  that if $C=AB$, then
\begin{align*}
\cc 11&= I+IV-V+VII,\\
\cc 21&= II+IV,\\
\cc 12 &= III + V,\\
\cc 22 &= I+III-II+VI.
\end{align*}

Thus the above is an algorithm for multiplying two by two
matrices performing only seven multiplications.

\begin{remark} Strassen was   attempting to prove,
by process of elimination, that such an 
algorithm did not exist when he arrived at it. We will see in \S\ref{geometricformulationsect} why the
result could have been anticipated using elementary
algebraic geometry.  
\end{remark}

\subsection{The complexity of matrix multiplication}\label{1.2}
In Strassen's algorithm  the entries of the matrices need not
be scalars - they could be elements of an algebra.
Let $A,B$ be $4\times 4$ matrices, and write
$$
A=\begin{pmatrix}a^1_1 & a^1_2 \\ a^2_1
&a^2_2\end{pmatrix},\ \
B=\begin{pmatrix}b^1_1 & b^1_2 \\ b^2_1
&b^2_2\end{pmatrix}.
%,\ \ 
%C=\begin{pmatrix}c^1_1 & c^1_2 \\ c^2_1
%&c^2_2\end{pmatrix}
$$
where $a^i_j,b^i_j$ are $2\times 2$ matrices. We may
apply 
Strassen's algorithm to get the blocks of $C=AB$ in terms
of the blocks of $A,B$ performing $7$ multiplications
of $2\times 2$ matrices. Since we can apply Strassen's algorithm
to each block, we can multiply $4\times 4$ matrices using
$7^2=49$ multiplications instead of the usual $4^3=64$.
In fact, if $A,B$ are $2^k\times 2^k$ matrices, we may multiply
them using $7^k$ multiplications rather than the usual $(2^{k})^3$.
Even if $n$ is not a power of two, we can still save multiplications
asymptotically by enlarging the dimensions of our matrices,
placing zeros in the new entries, to obtain matrices whose size is a power of two. Asymptotically  we can
multiply $n\times n$ matrices using
  $O(n^{log_2 (7)})\simeq O(n^{2.81})$ operations, as let $n=2^k$ and write $7^k=(2^k)^a$ so
$k(log_27)=ak(log_22)$ and we obtain $a=log_27$.

\begin{definition} The {\it exponent}\label{mexponentdef}  $\o$ of matrix multiplication is
%$$
%\o :=\tinf \{ h\in \BR \mid  Mat_{n\times n} 
%{\rm \  may\ be\ multiplied\ using\ } O(n^h) {\rm \ arithmetic\  %operations}
%\}
%$$
%In fact (see \cite{BCS}, \S 15.1) we can ignore additions and %subtractions:
$$
\o  =\tinf \{ h\in \BR \mid  Mat_{n\times n} 
{\rm \  may\ be\ multiplied\ using\ } O(n^h) {\rm \ scalar\  multiplications}
\}.
$$
\end{definition}
Strassen's algorithm shows $\o\leq log_2 (7) <2.81$.

\begin{remark}
If one replaces
the phrase \lq\lq scalar multiplications\rq\rq\
with the phrase \lq\lq arithmetic operations\rq\rq\
in the definition, $\o$ is unchanged, see \cite{BCS}, Proposition 15.1.
\end{remark}

\smallskip
 
Matrix multiplication of square matrices is a bilinear map
that we denote $M_{n,n,n}: \BC^{n^2}\times \BC^{n^2}
\ra \BC^{n^2}$. (In this article we restrict our
attention to the complex numbers, so e.g., all vector
spaces are finite dimensional vector spaces over $\BC$.)
When discussing a minimal number of arithmetic
operations (or multiplications) for executing a
bilinear map, it is usually within the context
of a class of algorithms. A natural class of algorithms for
executing a bilinear map is as follows: let $A,B,C$ be
vector spaces, let $A^*:=\{ f: A\ra \BC\mid f{\rm \ is\ linear}\}$ denote the dual
vector space (and similarly for $B$), and let
  $T: A\times B\ra  C$ be a bilinear map. Choose $\a^i\in A^*$,
$\b^i\in B^*$, $c_i\in C$ such that
$T(v,w)=\sum_{i=1}^r\a^i(v)\b^i(w)c_i$. The minimal number 
$r$ over all such presentations of $T$ is called the {\it rank} of $T$ and denoted $\trankc (T)$.
A related notion, more
natural to geometry and  defined in \S\ref{tensorformulation}, is
that of {\it border rank}, denoted $\tbrank (T)$.
  Another concept that comes into play when
discussing the space of all bilinear maps $A\times B\ra C$,
is the {\it typical rank}, which is the rank of
a generic bilinear map $A\times B\ra C$.

Strassen's algorithm shows that the rank of the multiplication
of two by two matrices is at most seven, and
  Winograd \cite{win553} proved that in fact it equals seven.

\subsection{Overview} To examine
the complexity of matrix multiplication more geometrically,
we first, in  \S\ref{tensorformulation},   rephrase it using
tensors.
Next,  in \S\ref{geometricformulationsect},
 we introduce algebraic varieties
which stratify the space of tensors, the {\it secant
varieties of Segre varieties}. (The above-mentioned
  border rank  of a tensor describes its location
with respect to this stratification.)
This is done in two steps,
first introducing secant varieties to any algebraic variety
in \S\ref{secantvarsect};
 then specializing to
Segre varieties in \S\ref{Tepexplan}. We also rephrase the main
open problems in the complexity of matrix multiplication
in terms of secant varieties of Segre varieties.
In \S\ref{updatesect} we summarize the known
results.

Before discussing those results in detail, we take two detours.
In the first,  we describe  two problems
from algebraic geometry where secant varieties arise:
the {\it polynomial Waring problem} and 
{\it Hartshorne's conjecture
on linear normality}. These are described in  
in \S\ref{secinalggeomsect}.
In the second, we
describe  other applications of secant varieties of
Segre varieties - to {\it algebraic statistics} (especially
the study of {\it phylogenetic invariants}) and {\it quantum
computing}, which is done in \S\ref{otherusessect}.
These detours will allow the reader to place the
topics discussed in the remainder of the paper in a larger
mathematical context.

In \S\ref{strasseneqnsect} we describe {\it Strassen's equations}
for secant varieties of Segre varieties
and their use in proving lower bounds for rank and
border rank.  In particular, we present a new
proof of {\it Bl\"asser's $\frac 52$-Theorem}. We  
rephrase Strassen's equations   invariantly
in \S\ref{invarstrassen} and describe  generalizations.

While it is well known that the limit
of a family of secant lines is a tangent line
(or a secant line itself), exactly what can
be in the limit of a secant $k$-plane is {\it not}
known. We discuss what is known about this problem in \S\ref{seclimitsect}
and show how to use this knowledge  
to prove upper bounds for the complexity of matrix
multiplication in \S\ref{Schonsect}. (We  
explain how to use   such limits to prove lower bounds
in the discussion below Theorem \ref{lmatrixthm}.)  A  
group-theoretic approach to upper bounds is described
briefly in \S\ref{finitegpsect}. 

We     discuss dimensions of secant
varieties of Segre varieties in \S\ref{secdimsect},
focusing on the use of {\it Terracini's Lemma}.

Any proper study of   varieties invariant under a group
action, e.g.,  the secant varieties of Segre varieties,
should exploit {\it representation theory}.
The representation theory relevant to
this study   is
discussed in
\S\ref{reptheorysect}.  
Representation theory is
 the most important tool discussed in this article.

 A common technique in
geometry is to understand a complicated geometric
object via the construction of auxiliary objects
that are more tractable, and the problem at
hand is no exception. We describe two such
objects in \S\ref{auxvarsect}.

  In 
\S\ref{weymansect}, we describe a   collection
of techniques developed by Weyman for the study
of $G$-varieties and their application  to secant
varieties of Segre varieties. (A $G$-variety is
a variety invariant under the action of an algebraic
group $G$.) These techniques  find the entire minimal free resolution
of the ideal of a variety 
and describe the nature of its singularities.

 Finally, in an appendix \S\ref{appendix}, we give nontraditional
and more invariant presentations
of two standard notions in complexity theory -
{\it multiplicative complexity} and {\it separations}.

%\subsection{Notation} Unless otherwise stated,
%$A,B,C,A_j,V,V_j$ denote finite dimensional complex vector
%spaces  with $A,B,C$ respectively of dimensions
%$\aaa,\bbb,\ccc$. (Although applications to finite fields are important,
%we do not deal with them in this article.)

\subsection{Acknowledgments} Many colleagues
generously helped the author in the preparation of
this article. Special thanks are due to
E. Allman, M. Bl\"aser,
P. B\"urgisser,  L. Garcia, D. Gross, J. Morton, G. Ottaviani, C. Robles
and the anonymous referee for numerous
suggestions to improve  this article.
In particular, the new proof of Bl\"aser's theorem
arose out of discussions with P. B\"urgisser.

\section{Tensor formulation}\label{tensorformulation}

Recall that for   vector spaces $V, V_j$,
\begin{align*}
V^*:&=\{ f: V\ra \BC \mid f{\rm \ is \ linear}\},\\
V_1\otc V_n:&=\{ f: V_1^*\ctimes V_n^*\ra \BC
\mid f{\rm\ is\ linear \ in \ each \ factor}\}.
\end{align*}
Given $v_j\in V_j$, $\a_j\in V_j^*$,   define
$v_1\otc v_n\in V_1\otc V_n$ by 
$v_1\otc v_n(\a_1\hd \a_n)=\a_1(v_1)\cdots \a_n(v_n)$.
An element $f\in V_1\ot V_2$, i.e., a bilinear map
$f:V_1^*\times V_2^*\ra\BR$, may also be considered
as a linear map 
\begin{align*}
f: V_1^*&\ra V_2\\
\a&\mapsto f(\a,\cdot)
\end{align*}
where  $f(\a,\cdot )\in (V_2^*)^*= V_2$, i.e., for  $\b\in V_2^*$,
$f(\a,\cdot )(\b)= f(\a,\b)$.

\begin{definition}\label{rankdef} Let $V_1\hd V_k$ be vector spaces. An element $z\in V_1\otc V_k$ is
called {\it decomposable} if there exist $v_i\in V_i$ such that
$z=v_1\otc v_k$. Define the {\it rank}\index{rank!of a tensor} of an element
$T\in V_1 \ot V_2 \ot\hdots\ot V_k $ to be the minimal number
$r$ such that $T=\sum_{u=1}^rz_u$
with each $z_u$ decomposable. We refer to an explicit
expression for a tensor  $T$ as a sum of $r$ monomials
as a {\it computation of $T$ of length $r$}, and sometimes
use   $\phi$ to denote the realization
of $T$ as a computation. This terminology
is   consistent with the definition of the rank
of a linear map $T: V_1^*\ra V_2$ (i.e., an element   $T\in V_1\ot V_2$)
and the rank of a bilinear  map $T: V_1^*\times V_2^*\ra V_3$ given in \S\ref{1.2}  
(i.e., an element of $T\in V_1\ot V_2\ot V_3=A^*\ot B^*\ot C$).
Note  that the length of a computation of a tensor
is unchanged if we make changes of bases in the vector spaces
$V_i$. 
\end{definition}

\subsection{Strassen's algorithm as a tensor}
The standard algorithm for the multiplication
of two by two matrices in terms of tensors as
follows: let $A,B,C$ each denote the space of
$2\times 2$ matrices; give $A$ the standard basis $a^i_j$ for
the matrix with a $1$  in the $(i,j)$-th slot and zeros elsewhere,
and let $\a^i_j$ denote the corresponding
elements of the  dual basis of $A^*$. Similarly for
$B,C$. Then the standard algorithm is (compare with \eqref{stdalg}:
\def\aa#1#2{\a^{#1}_{#2}}\def\bb#1#2{\b^{#1}_{#2}}\def\cc#1#2{c^{#1}_{#2}}
\begin{equation}\label{standardalg}
\begin{aligned} 
M_{2,2,2}=&
\aa 11\ot \bb 11\ot \cc 11
+\aa 12\ot\bb 21\ot \cc 11+\aa 21\ot\bb 11\ot\cc 21+\aa 22\ot\bb 21\ot\cc 21
\\
&
+
\aa 11\ot \bb 12\ot \cc 12
+\aa 12\ot\bb 22\ot \cc 12+\aa 21\ot\bb 12\ot\cc 22+\aa 22\ot\bb 22\ot\cc 22
\end{aligned}
\end{equation}
and Strassen's algorithm is
\begin{equation}\label{strassen69}
\begin{aligned}
  M_{2,2,2}=&   (\aa 11 + \aa 22)\ot(\bb 11 + \bb 22)\ot (\cc 11+\cc 22)
+(\aa 21 + \aa 22)\ot \bb 11\ot (\cc 21 -\cc 22)
\\
&
+\aa 11\ot (\bb12-\bb 22)\ot (\cc 12 +\cc 22) +\aa 22\ot(-\bb 11+\bb 21)\ot (\cc 21 +\cc 11)\\
&
+(\aa 11+\aa 12)\ot \bb 22\ot (-\cc 11+\cc 12)
+(-\aa 11+\aa 21)\ot (\bb 11 +\bb 12)\ot \cc 22\\
&
+(\aa 12 -\aa 22)\ot (\bb 21 + \bb 22)\ot \cc 11.
\end{aligned}
\end{equation}

\subsection{Approximate algorithms}
An  {\it approximate algorithm}
for a tensor $T$  is a   sequence  of algorithms, usually
 of lower rank tensors,  that converge  to  
an algorithm for $T$. The {\it border rank} of a tensor
$T$
is the lowest rank of tensors in such   sequences 
and is denoted $\tbrank (T)$.
 Note that rank and border rank can  indeed  be different - consider
the following example:

\begin{equation}\label{Ttanten}
T=a_1\ot b_1\ot c_1 + a_1\ot b_1\ot c_2 + a_1\ot b_2\ot c_1 
+a_2\ot b_1\ot c_1
\end{equation}
One can show that   $\trankc(T)=3$, but we can
approximate $T$ as closely as we like by tensors of rank two
as follows. Let
\begin{equation}\label{Tep}
T(\ep ) =\frac 1\ep[(\ep -1) a_1\ot b_1\ot c_1 + (a_1+\ep a_2)\ot (b_1+\ep b_2)
\ot (c_1+\ep c_2)]
\end{equation} 
and allow $\ep\ra 0$, so $\tbrank(T)\leq 2$ (in fact equality holds).
The geometry of this limit is discussed in \S \ref{Tepexplan}.

 \section{Geometric formulation}\label{geometricformulationsect}

\subsection{Secant varieties}\label{secantvarsect}
Let $V$ be vector space and let $\BP V$ be the associated projective space of lines
through the origin in $V$, so we have a map
$\pi:V\backslash 0\ra \BP V$.
If $v\in V\backslash 0$,   let $[v]=\pi(v)\in \BP V$
and
for $Z\subset \BP V$,   let $\hat Z=\pi\inv(Z)\subset V$.
For scale invariant sets $U\subset V\backslash 0$, 
  write $\BP U$ for $\pi (U)$. 
We use projective space in addition to vector spaces
because
the properties
we are interested in (rank, border rank) are scale
invariant.
Because we go back and forth between vector
and projective spaces 
  many objects end up being decorated with hats and \lq\lq$\BP$\rq\rq s

 For our purposes, a {\it variety}
$X\subset \BP V$ is
the common zero locus in $\BP V$ of a collection of homogeneous polynomials
on $V$. Given a variety $X$, we will construct a sequence
of auxiliary varieties $X\subset \s_2(X)\subset
\cdots \subset \s_f(X)= \BP V$, called
the {\it secant varieties of $X$} which 
determine a stratification of $\BP V$. 
This stratification will generalize the stratification
of the space of $m\times n$ matrices by rank. 
  When $V=A_1\otc A_n$
and $X$ is the projectivization of
the set of decomposable tensors, the stratification
will  coincide   with the stratification of tensors by their border rank,
and $f$ is the {\it typical rank}
mentioned in \S\ref{section1} and defined below.

\smallskip

For   readers not accustomed to 
secant varieties, we    begin with several
special cases to help   visualize them.
Recall that projective space $\BP V$ has the property that,
given any two distinct points $p,q\in \BP V$, there is a unique
{\it line}, i.e., a linearly embedded $\pp 1\subset \BP V$ 
containing $p$ and $q$, which we denote $\BP^1_{p,q}$.
Let $C\subset \BP V$ be a smooth curve (one-dimensional variety) and $q\in \BP V$ a point.
Let $J(q,C)\subset \BP V$ denote the {\it cone}
over $C$ with vertex $q$, which
by definition contains the 
  union of all points on
all lines containing $q$ and a point of $C$. More
precisely,  $J(q,C)$ denotes the  closure
of the set of such points. It  is only necessary
to take the closure when   $q\in C$, as in this case
one also includes the points on the tangent line to $C$ at $q$, because,
as anyone who has ever taught calculus knows,
the tangent line is the
limit of secant lines $\BP^1_{q,x_j}$ as $x_j\ra q$.
Define  $J(q,Z)$ similarly  for $Z\subset \BP V$,  a
variety of any dimension. Unless $Z$ is a linear space
and $q\in Z$,   $\tdim J(q,Z)=\tdim Z +1$.

\begin{definition}\label{joinetcdefs}
The {\it join}\index{join of varieties} of $Y,Z\subset \BP V$
is
$$
J(Y, Z)=\overline{\textstyle{\bigcup_{x\in Y, y\in Z, x\neq y}\pp 1_{xy}} }.
\index{$\mrom{J}J(Y,Z)$, join of varieties}
$$
Here   the overline denotes Zariski closure, i.e., if $U\subset \BP V$
is a subset,
then $\overline U$ is the common zero set of all homogeneous polynomials
vanishing on $U$. The same set is obtained if one
takes the closure in the usual topology, but the
Zariski closure is more useful when
dealing with polynomials. 
We may   think of 
$J(Y,Z)$ as the union of the cones $\cup_{q\in Y}J(q,Z)$
(or as the union of the cones over $Y$ with vertices
points of $Z$.)

  If $Y=Z$,
we call $\sigma_2 (Y)= J(Y,Y)$ the {\it secant variety}
\index{secant variety}
\index{variety!secant} of $Y$. By the discussion above,
$\s_2(Y)$ contains all points of all secant
and tangent lines to
$Y$.
Similarly, define the join of $k$ varieties to be the
closure of the 
union of the corresponding $\pp{k-1}$'s, or by induction as
$J(Y_1\hd Y_k)=J(Y_1,J(Y_2\hd Y_k))$.
Define {\it $k$-th secant variety of $Y$}
to be  $\s_k(Y)=J(Y\hd Y)$,  the join
of $k$ copies of $Y$.
For smooth varieties $Y\subset \BP V$, let $\t(Y)$
denote the union of all points on all embedded tangent
lines to $Y$. Usually $\t(Y)$ is a hypersurface in $\s(Y)$.
\end{definition}

\begin{remark}\label{secdimexpect}
The expected   dimension of
$J(Y,Z)$ is 
$\tmin\{ \tdim Y+\tdim Z+1, \tdim \BP V\}$  because a point
  $x\in J(Y,Z)$ is obtained by picking a point of $Y$, a point
of $Z$, and a point on the line joining the two points.
This expectation fails if and only if  a general point of $J(Y,Z)$ lies
on a family of lines intersecting $Y$ and $Z$, as when this happens
one can vary the points on $Y$ and $Z$ used to form the
secant line without varying the point $x$.

Similarly, the expected dimension of $ \s_r(Y)$
is $r(\tdim Y)+r-1$ which fails
 if and only if  a general point of $\s_r(Y)$ lies
on a family of secant $\pp{r-1}$'s to $Y$.
\end{remark}

 \begin{definition}
For a    variety $X\subset  \BP V$, and
point  $p\in \BP V$,
  the {\it $X$-rank of $p$} is the smallest number $r$ such that $p$ is in the linear
span of $r$ points of $X$. Thus
$\s_r(X)$ is the Zariski closure of the set of points of $X$-rank $r$.
The {\it $X$-border rank of $p$} is the smallest $r$ such that
$p\in \s_r(X)$. The {\it typical $X$-rank}  of $\BP V$ is
the smallest $r$ such that $\s_r(X)=\BP V$.
\end{definition}

\subsection{The Segre variety and border rank}\label{Tepexplan}
Define $Seg(\BP V_1\times \BP V_2)\subset \BP (V_1\ot V_2)$,
the {\it (two-factor) Segre variety} to be the projectivization
of all the rank one elements of $V_1\ot V_2$.
Here   $Seg$ is the injective map
\begin{align*}
Seg: \BP V_1\times \BP V_2& \ra \BP (V_1\ot V_2)\\
([v_1],[v_1])&\mapsto [v_1\otimes v_2]
\end{align*}
which, in bases, corresponds to multiplying a 
column vector (defined
up to scale) with a  row vector (defined
up to scale) to get a  rank one  rectangular
matrix (defined
up to scale). Note that
$\hat\s_r(Seg(\BP V_1\times \BP V_2))$ is isomorphic
to the set of $(\tdim V_1\times \tdim V_2)$ matrices
of rank at most $r$, as the rank at most $r$ matrices
are exactly those that can be written as the sum of
$r$ matrices of rank one.

More generally, the projectivization of the 
set of decomposable tensors in $V_1\otc V_n$,
i.e., $\BP \{T\in V_1\otc V_n\mid \exists v_j\in V_j,
\ T=v_1\otc v_n\}$, may be identified
with the product $\BP V_1\ctimes \BP V_n$.
Let $Seg(\BP V_1\ctimes \BP V_n)\subset \BP(V_1\otc V_n)$
denote the corresponding variety,
  the {\it ($n$-factor) Segre variety}.

For any variety $X$, a point of $\s_2(X)$ is a point
on a limit of secant lines, so if $X$ is smooth, the point
is either on $X$, on a  secant line, or on a tangent
line to $X$. Equation \eqref{Tep},
when projectivized,  is a curve of
points on secant
lines of $Seg(\pp 1\times\pp 1\times \pp 1)$ limiting to a point on a tangent line
to $Seg(\pp 1\times\pp 1\times \pp 1)$, i.e., 
    a
  point
of $\hat\t(Seg(\BP A\times \BP B\times \BP C))$.

We can now give  geometric formulations of the
concepts introduced in \S\ref{section1} and \S\ref{tensorformulation}:

\medskip

\begin{itemize}

\item  The {\it border rank}  of
a tensor $T\in V_1\otc V_n$,
$\tbrank (T)$,    defined  in  \S\ref{rankdef} above,
is the smallest $r$ such that $[T]\in \s_r(Seg(\BP V_1\ctimes\BP V_n))$.

\item The {\it border rank of matrix multiplication} 
$$M_{m,n,p}:
(\BC^{m*}\ot \BC^n)\times (\BC^{n*}\ot \BC^p)\ra
(\BC^{m*}\ot \BC^p)
$$ 
is the
smallest $r$ such that 
$$[M_{m,n,p}]\in \s_r(
Seg(
\BP(\BC^{m}\ot \BC^{n*})\times \BP (\BC^{n}\ot \BC^{p*})
\times \BP
(\BC^{m*}\ot \BC^p)))
$$

\item
The {\it exponent of matrix multiplication} is 
$$ 
\tliminf_{n\ra\infty}\{\tmin_r\{
[M_{n,n,n}]\in \s_r(Seg(\BP^{n^2-1}\times \BP^{n^2-1}
\times \BP^{n^2-1})\}\}
$$

\item
{\it Upper   bounds for border rank}
for a given $n$ can be proven
by finding values of $r$ such that
$[M_{n,n,n}]\in \s_r(Seg(\BP^{n^2-1}\times \BP^{n^2-1}
\times \BP^{n^2-1}))$ 
and lower bounds by finding values of $r$ such that
$[M_{n,n,n}]\notin \s_r(Seg(\BP^{n^2-1}\times \BP^{n^2-1}
\times \BP^{n^2-1}))$.

\item The {\it typical rank} of an element of $\BC^a\ot \BC^b\ot \BC^c$
is the smallest $r$ such that $\s_r(Seg(\BP^{a-1}\times
\BP^{b-1}\times \BP^{c-1}))=\BP(\BC^a\ot \BC^b\ot \BC^c)$.
\end{itemize}

\medskip

\subsection{What is known regarding matrix multiplication}\label{updatesect}
The problem of determining the typical rank for
the spaces   that include the multiplication
of square matrices has been completely solved:

\begin{theorem}[Lickteig \cite{lick}]\label{licksecdims} For all $n\neq 3$,
$$
\tdim \s_r(Seg(\pp{n-1}\times \pp{n-1}\times \pp{n-1}))
=\tmin\{ r(3n-2) -1, n^3-1\}.
$$
\end{theorem}

In particular note that Theorem \ref{licksecdims}
shows that Strassen's algorithm for $2\times 2$ matrices
could have been anticipated, as $\s_7(Seg(\pp 3\times \pp 3\times\pp 3))
=\BP (\BC^4\ot\BC^4\ot\BC^4)$.
We outline the proof of Theorem
\ref{licksecdims}  and discuss what is known
  about typical rank in \S\ref{secdimsect}.

For the $n=3$ case, we have:

\begin{theorem}[Strassen, \cite{strassen505}]\label{thm6.1} $\s_4(Seg(\pp 2\times \pp 2\times \pp 2))$ is a hypersurface
of degree nine.
\end{theorem}

This case was solved by
finding an explicit equation vanishing on $\s_4(Seg(\pp 2\times \pp 2\times \pp 2))$.
In \S\ref{strasseneqnsect} we discuss this equation and its consequences
for matrix multiplication.

\smallskip

\medskip

The best   lower bound on the border rank
of matrix multiplication is:

\begin{theorem}[Lickteig \cite{MR86c:68040}]
 $\tbrank (M_{m,m,m})\geq \frac {3m^2}2+\frac m2 -1$.
\end{theorem}

While we do not provide Lickteig's proof here, we
remark that implicit in his proof are the presence
of auxiliary varieties which we believe will play
a central role in future work. \S\ref{auxvarsect}
describes some of these varieties, including the
{\it subspace variety} that is implicit in his proof.

\medskip

The best   lower bounds on the rank
of matrix multiplication are:

\begin{theorem}[Bl\"aser \cite{Bl1}]\label{Blaser52}
$\trankc (M_{m,m,m})\geq\frac 52 m^2-3m$.
\end{theorem}

A new proof of Bl\"aser's theorem is presented in \S\ref{blaserpf}.
Bl\"aser has also proved that 
$\trankc (M_{3,3,3})\geq 19$ \cite{Bl2},
and we discuss the main tool in the proof of
Bl\"aser's $19$-theorem in \S\ref{separationsect}.

The best   upper bound for 
the exponent of matrix multiplication is
$\o <2.38$ due to Coppersmith and Winograd
\cite{copwin135}. They use methods of
Strassen \cite{MR882307}. We do not discuss these
asymptotic bounds as we have
no geometric interpretation for them.
However,  an earlier asymptotic
bound due to
Sch\"onhage \cite{schon460} does have relations with
geometry. We discuss the geometric aspect
of Sch\"onhage's argument in \S\ref{Schonsect}, 
and present his explicit
approximate algorithm for multiplying three by three
matrices using 21 multiplications. 

 There is also an algorithm for multiplying 
$3\times 3$ matrices using $23$ multiplications
due to Laderman \cite{Laderman} which we do not discuss.

\medskip

The only case where the exact rank and border rank
are known for the multiplication of square matrices
are two by two matrices:

\begin{theorem}[Winograd \cite{win553}]\label{rk2by2}
${\bold  R}(M_{2,2,2})=7$.
\end{theorem}

Hopcroft and Kerr \cite{hk252} proved Theorem \ref{rk2by2}
in the case of algorithms with integer coefficients.
 
While we do not discuss the original proof of Theorem \ref{rk2by2},
an alternative proof   is   a consequence
of a theorem of Brockett and Dobkin \cite{BD}  
 that the rank of the multiplication in any simple algebra
is at least twice the dimension of the algebra minus one.
A proof of the Brockett-Dobkin theorem, due
to Baur and presented in \cite{BCS}, proceeds by  splitting any putative simpler
algorithm  several times to eventually obtain a contradiction
by producing a right ideal that is contained in a left ideal.

\begin{theorem}[\cite{Lmatrix}]\label{lmatrixthm}
$\underline{\bold R}(M_{2,2,2})=7$.
\end{theorem}

To prove Theorem \ref{lmatrixthm}  we first decomposed 
$\s_6(Seg(\pp 3\times \pp 3\times\pp 3))$
into various components based on how the limiting $\pp 5$
was obtained from family of secant $\pp 5$'s. (By Theorem
\ref{rk2by2} one only needs to examine limiting planes.)
For each possible
limiting type we wrote down normal
forms for the limit. Then
we applied variants of Baur's proof of the Brockett-Dobkin
theorem in each case to obtain a contradiction.
In \S\ref{seclimitsect} we give an idea how to study
such limiting planes, which is also used in
the construction of upper bounds.

\subsection{What is not known}
The central conjecture
in algebraic complexity theory is that the exponent of matrix multiplication
is two. It is also of importance to find good upper and lower bounds for
matrix multiplication for
small and human scale values of $n$.
Already for $n=3$ all that is known is
$14\leq \underline{\trankc}(M_{3,3,3})\leq 21$, and $19\leq {\bold R}(M_{3,3,3})\leq 23$.
While the problem of finding the defining equations for
secant varieties of Segre varieties is   
a means to an end as far as matrix multiplication is
concerned, for the purposes of algebraic statistics,
it is essential to develop techniques for finding these
equations and the equations of related
varieties. For the area of phylogenetic invariants,
an important open problem is to find the defining
equations for $\s_4(Seg(\pp 3\times \pp 3\times\pp 3))$
as explained in \S\ref{phylo}. Other open questions are
discussed in the remaining sections.

\section{Secant varieties in algebraic geometry}\label{secinalggeomsect}

In this section we take a detour from our main subject
to discuss other situations where secant varieties
arise: the solution of the polynomial
Waring problem and the resolution of Hartshorne's
conjecture on linear normality.

\subsection{The   Waring
problem for polynomials and variants}\label{waring}
The Waring problem for polynomials is as follows:

\smallskip

{\it  What is 
the smallest $r_0=r_0(d,n)$ such that a
general   homogeneous polynomial $P(x^1\hd x^n)$ of
degree $d$ in $n$ variables
 is expressable
as the sum of $r_0$ $d$-th powers of linear forms?}

\smallskip

Let $V=\BC^n$, and let $S^dV^*$ denote the space
of homogeneous polynomials of degree $d$ on $V$.
Let 
\begin{align*}
v_d: \BP V^*&\ra \BP S^dV^*\\
[\a]&\mapsto [\a\circ\cdots \circ \a]
\end{align*}
denote the {\it Veronese map} that sends
the projectivization of
a linear form to the projectivization of its
$d$-th power. Thus the image is
the set of (projectivized) $d$-th powers of linear forms.
Similary  $\s_p(v_d(\BP V))$ is the Zariski
closure of the set of homogeneous polynomials
that are expressable as the sum of $p$ $d$-th
powers of linear forms. So the Waring problem
for polynomials may be re-expressed as:

\smallskip

{\it Let $V=\BC^n$ and let $X=v_d(\BP V^*)$.
What is 
the typical $X$-rank of an element of $\BP S^dV^*$,
i.e., what is the smallest $r_0=r_0(d,n)$ such that
$\s_{r_0}(v_d(\BP V^*))=\BP S^dV^*$?
}

\smallskip

 This problem was   solved by
Alexander and Hirshowitz  \cite{AH}:
all $\s_r(v_d(\pp n))$    are of the 
expected dimension except $\s_7(v_3(\pp 4)),\s_5(v_4(\pp 2)),  \s_9(v_4(\pp 3)), \s_{14}(v_4(\pp 4))$, (which are all hypersurfaces),
and $\s_r(v_2(\pp n))$, $2\leq r\leq n$
(where $\tdim\s_r(v_2(\pp n))=rn-\frac{r^2-3r}2-1$) . In other words,

\begin{theorem}\cite{AH}
A general homogeneous polynomial of degree $d$ in
$n$ variables is expressable as the sum of
$$
r_0(d,n)=\ulcorner   \frac { \binom{n+d-1}d +1  }n \urcorner
$$
$d$-th powers
with the exception of the cases
 $r_0(3,5)=8$, $r_0(4,3)=6$,
$r_0(4,4)=10$,   $r_0(4,5)=15$, and $d=2$, where $r_0(2,n)=n$.
\end{theorem}

For a beautiful discussion of this problem
and its history, including  a self-contained proof,
see \cite{Ottwaring}.

\smallskip

A  variant of the polynomial Waring problem
is to find the typical rank of alternating tensors.
Let $\La k V\subset V^{\ot k}$ be the space
of alternating tensors. Let $G(k,V)\subset \BP(\La k V)$
denote the projectivization  of the set of minimal rank
alternating tensors. This variety is called
the {\it Grassmanian} of $k$-planes through the origin
in $V$  (i.e.,
we have a bijection, for linearly independent
sets of vectors $v_1\hd v_k$,
$\tspan\{ v_k\hd v_k\}\simeq [v_1\ww\cdots\ww v_k]$).
In  \cite{MR2113908}   they show
that for
$3\le k\le\frac{n}2$, $\s_r(G(k,n))$ has 
the expected dimension provided that $r\le\frac{n}{k}$.  
Previous to that, it was known that
$G(2,n)$ had all secant varieties defective and  
$G(3,7)$, $G(4,8)$, and $G(3,9)$ all had
their \lq\lq last\rq\rq\ secant variety before
filling defective. (The examples $G(2,n)$ are just
the skew symmetric matrices of minimal rank; the examples  
$G(3,7)$ and $G(4,8)$ can be understood
in terms of the geometry of the exceptional
groups $G_2$ and $Spin_7$.)

Further generalizations of the polynomial Waring
problem and their uses are discussed in 
\cite{ciro}.

The main tool for proving secant varieties are of
the expected dimension is {\it Terracini's Lemma} 
  \ref{Terracinilemma}. Proving they
are degenerate, other than in cases when it is
obvious, is more subtle.
For all the  Waring problems, there appears
to be interpretations of the exceptional cases
in terms of the geometry of   Veronese varieties.
The most interesting exception in the case of secant varieties
of Segre varieties is $\s_4(Seg(\pp 2\times \pp 2\times \pp 2))$
which is discussed in detail in \S\ref{strasseneqnsect}.
In  the proof of Lemma 3.16  of \cite{AOP}, a geometric
explanation of the degeneracy is given: any four points
on $Seg(\pp 2\times\pp 2\times \pp 2)$ lie in some
$v_3(\pp 2)\subset \BP (S^3\BC^2)\subset \BP (\BC^2\ot\BC^2\ot \BC^2)$.
Thus when one applies Terracini's lemma, each of the four 
embedded tangent
spaces to the Segre must have at least a two-dimensional
subspace in the $\BP (S^3\BC^2)=\pp 9$, forcing 
a degeneracy.
It would be interesting to have a systematic understanding
of the Veronese varieties that unirule these exceptional
cases, e.g., in terms of representation-theoretic data.

\subsection{Zak's theorems}
Smooth projective varieties $X^n\subset \pp\na$ of
small codimension were shown by
Barth and Larsen (see, e.g., \cite{Barthicm})
to behave topologically as if they were {\it complete interesections},
i.e, the zero set of $a$ homogeneous polynomials.
This motivated Hartshorne's famous conjecture
on complete intersections \cite{MR0384816}, which
says that if $a<\frac n2$, then $X$ must indeed be
a complete intersection. A first approximation
to this difficult conjecture was also made
by Hartshorne - his conjecture on {\it linear
normality}, which was proved by Zak \cite{Zakhart} (see \cite{zak}
for an exposition).
The linear normality conjecture was
equivalent to showing that
if $a<\frac n2+2$, and $X$ is not contained
in a hyperplane, then $\s_2(X)=\pp\na$.
Zak went on to classify the exceptions
in the equality case
$a=\frac n2+2$. There are exactly
four, which Zak called 
{\it Severi varieties} (after Severi, who
solved the $n=2$ case \cite{Severi}). The first three Severi
varieties have already been introduced:
$v_2(\pp 2)\subset \pp 5$,
$Seg(\pp 2\times \pp 2)\subset \pp 7$, and
$G(2,6)\subset \pp {13}$. The last is
the complexified Cayley plane
$\BO\BP^2\subset \BP^{15}$. These four varieties admit
a uniform interpretation as the rank
one elements in a rank three Jordan
algebra over a composition algebra.

An interesting open question is the {\it secant defect
problem}. For a smooth projective variety
$X^n\subset \BP V$,
not contained in a hyperplane, with
$\s_2(X)\neq \BP V$,  let $\d (X^n)=2n+1-\tdim \s_2(X)$,
the {\it secant defect of $X$}. The largest known
secant defect is $8$, which occurs for the complexified
Cayley plane. Problem: Is a larger secant defect
than $8$ possible? If we do not
assume the variety is smooth, the defect is unbounded.
(This question was posed originally in  \cite{LVdV}.)

\section{Other uses of secant varieties of Segre varieties
and related objects}\label{otherusessect}

\subsection{Algebraic Statistics}

  A {\it probability distribution} is a point in
$V:=\BR^{a_1}\otc \BR^{a_n}$ where the sums of coordinate elements add to one. For example,
say we have two biased coins. Then $V=\BR^2\ot \BR^2$ and a point corresponds
to a matrix
$$
\begin{pmatrix}
p_{h,h} &p_{h,t}\\ p_{t,h}&p_{t,t}\end{pmatrix}
$$
where $p_{h,h}$ is the probability that both coins, when tossed, come up
heads,  etc...

 A {\it statistical model} is a family of probability
distributions given by a set of contraints that these distributions must
satisfy, i.e., a subset of $V$. An {\it algebraic statistical model} consists of all joint
probability distributions that are the common zeros of a set of
polynomials on $V$.

To continue our example, assume the outcome of the two
coin tosses do not effect each other (as is the case with actual coins).
Then the resulting matrix must have rank one. The set of all rank one,
$2\times 2$   matrices in the positive coordinate simplex
is the corresponding algebraic statistical model, but
it is almost equivalent to work with
$\hat Seg(\BR\pp 1\times \BR\pp 1)$.

\smallskip
 
Now assume we   can   measure
the outcome of two of the
events  (tosses) but there may be a third event
whose outcome influences the outcome of the other two
although the outcomes of the two events we can measure are independent of one another  (e.g. someone may be cheating
by using magnets).

Na\"\i vely we should have a point of $\BR^{a_1}\ot\BR^{a_2}\ot \BR^{a_3}$
but   we can't measure the possible third,
in fact we don't even know what $a_3$ should be. 

Let's posit that some fixed $a_3$ parametrizes
the third outcome (if we posit there is no third event, then one
takes $a_3=1$). Then
 we sum up over all possibilities
for the third factor
to get a $2\times 2$ matrix whose entries are
\begin{equation}\label{star}
p_{i,j}= p_{i,j,1}+\cdots +p_{i,j,a_3},\ \ 1\leq i\leq a_1,\ 1\leq j\leq a_2
\end{equation}
The algebraic statistical model here is the set of rank at most $a_3$ matrices
in the space of $a_1\times a_2$ matrices,
$\hat\s_{a_3}(Seg(\BR\pp{a_1-1}\times \BR\pp{a_2-1}))$.
Thus, given a particular model, e.g.
a fixed value of $a_3$,
to test   if our   data (as points of $\BR^{a_1}\ot \BR^{a_2}$)
fits the model, we can check if it (mostly) 
lies inside $\hat \s_{a_3}(Seg(\BR\pp{a_1-1}\times \BR\pp{a_2-1}))$.

In algebraic statistics one wants to test if a given model is applicable to
a particular collection of data sets. Thus in particular, one needs a way
of testing if a point $p\in \BR^{a_1}\otc \BR^{a_n}$ is
a sum of at most $r$ decomposable elements.

It is easier to solve this problem
first over the complex numbers and then return
to the real situation later. Thus to test models
of the type discussed above,  one needs defining
equations for secant varieties of Segre varieties.
In sections \S \ref{strasseneqnsect} -   \ref{weymansect} we discuss methods for finding such
equations. These methods are applicable to finding equations
for more general algebraic statistical models
as well. They all rely on exploiting the group
  under which the model is invariant.

For more on algebraic statistics, see
\cite{MR2227048,MR2205865}.

\subsection{Phylogenetic invariants}\label{phylo}
This is   a special case of algebraic statistics, but
  is sufficiently important to merit its own subsection.
In order to determine a tree that describes the evolutionary
descent of a family of extant species, Lake \cite{Lake},
Cavender and Felsenstein \cite{cavenderfelstein} proposed
the use of what is now called algebraic statistics by
viewing the four bases composing DNA as the
possible outcomes of a random variable.

Given a collection of extant species, one would
like to assess the likelyhood of each of the possible
evolutionary trees that could have led to them.
To do this, one can test the various
DNA sequences that arise to see which   algebraic
statistical model  fits best. More than that, the
invariants discussed below identify the trees (nearly) uniquely.

In what follows, contrary to some of the literature,
we ignore time.

The simplest situation is where one species  
  gives rise to two new species. This can
be pictured by a tree of the form

\begin{figure}[!htb]\begin{center}
\includegraphics[scale=.3]{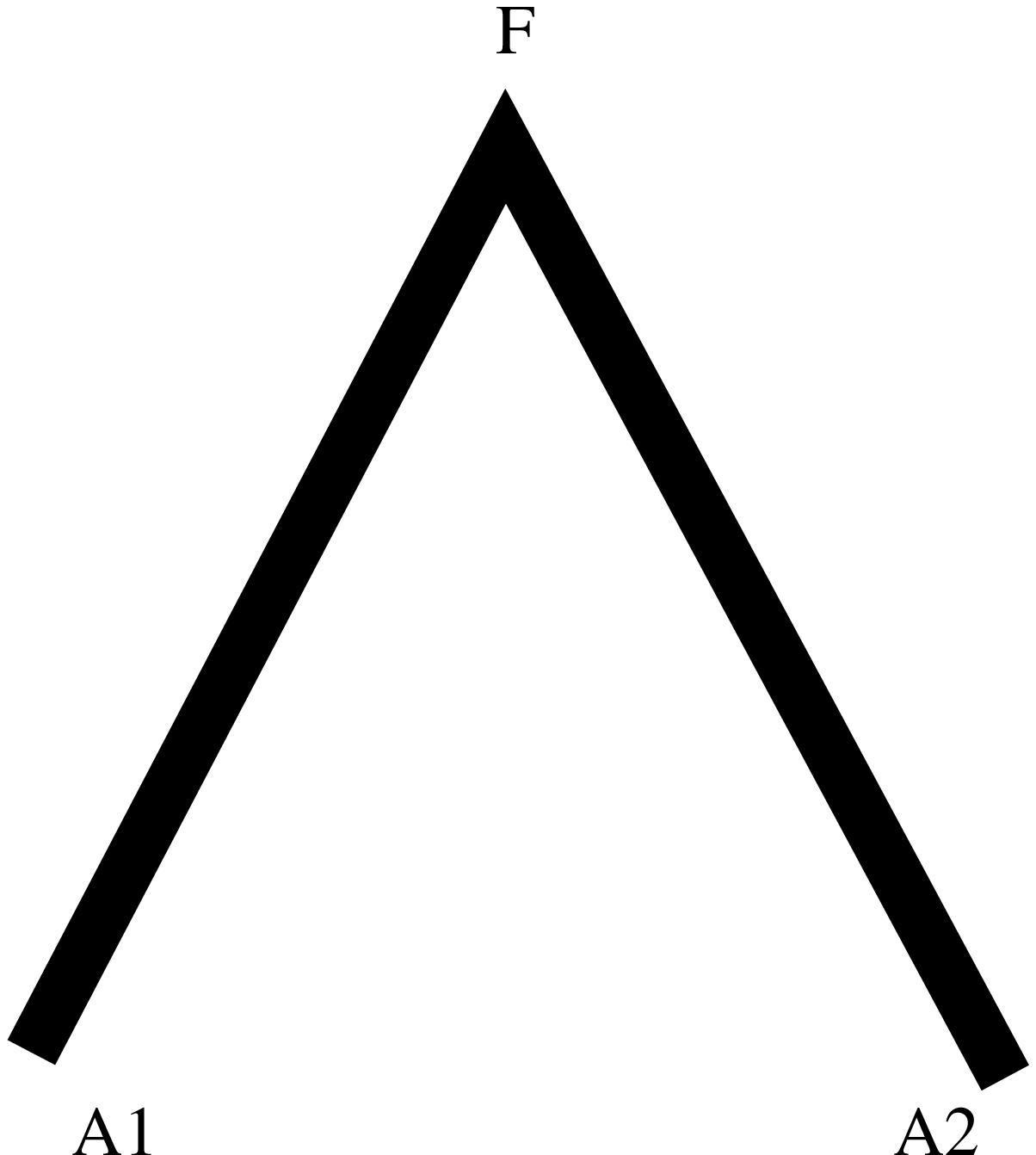}
\caption{\small{}}  
\end{center}
\end{figure}

There are three species involved, the parent $F$ and
the two offspring $A1,A2$, so the DNA occupies a point
of the positive coordinate simplex in  $\BR^4\ot \BR^4\ot \BR^4$, and  
we make our lives easier by working with
$\BP (\BC^4\ot \BC^4\ot \BC^4)$.
We can measure the DNA of the two new species but
not the ancestor, so the relevant
algebraic statistical model is
$\s_4(Seg(\BP^3\times \BP^3))$, which is well
understood.  Here $a_1=a_2=a_3 $ in the
analogue of equation \eqref{star} and we sum over the third factor.  
In this case there is nothing new
to be learned from the model.

\smallskip

The next case is where a parent $F$ gives rise to three
new species $A1,A2,A3$. Assuming species bifurcate,
one might think that this gives
rise to three distinct algebraic statistical
models, as we could have $F$ giving rise to $A_1$ and
$G$, then $G$ splitting to $A_2$ and $A_3$ or
two other possibilities.
However, all three senarios give rise to the
same algebraic statistical model: 
$\s_4(Seg(\BP^3\times \BP^3\times \BP^3))$.
(See \cite{AR2}.)
In other words, the following pictures all give rise to the 
same algebraic statistical models.

\begin{figure}[!htb]\begin{center}
\includegraphics[scale=.25]{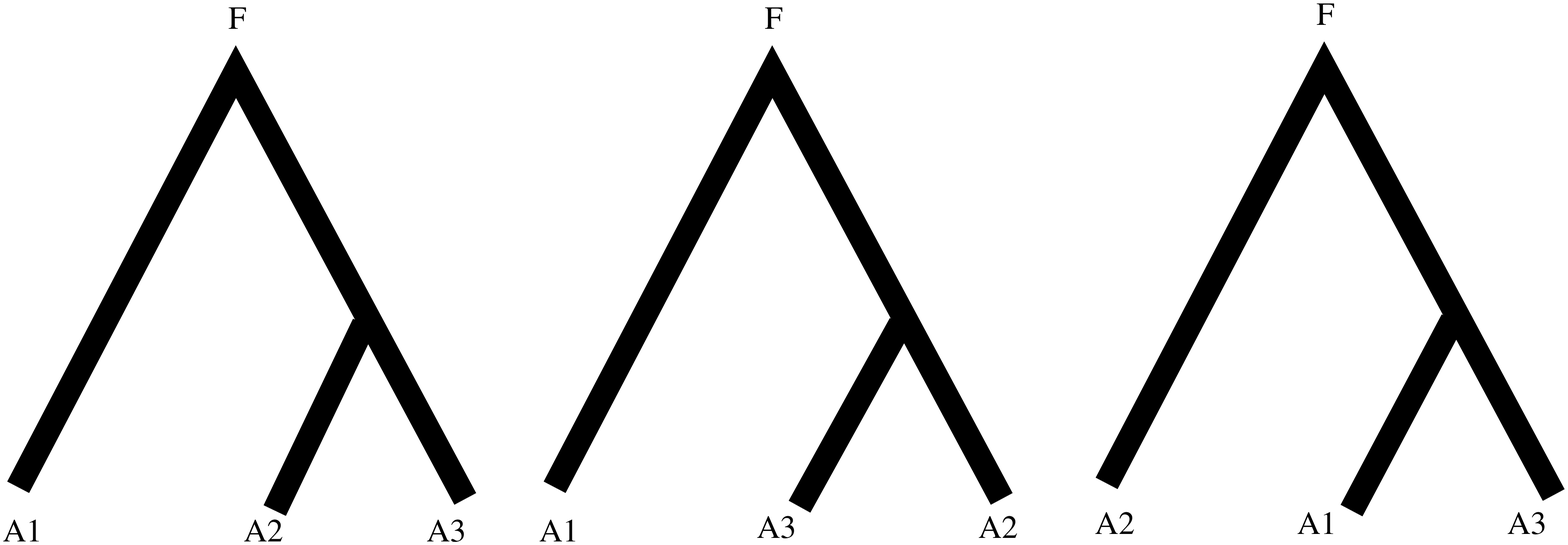}
\caption{\small{}}  
\end{center}
\end{figure}

The
defining equations of $\s_4(Seg(\BP^3\times \BP^3\times \BP^3))$ are {\it not known}, and for reasons we   explain below, it is
a central question for the study of phylogenetic
invariants to find them.

\smallskip

Now consider the case where there are four new species
$A1,A2,A3,A4$
all from a common ancestor $F$. Here finally there are
three different senarios  that give rise to distinct 
algebraic statistical models.

\begin{figure}[!htb]\begin{center}
\includegraphics[scale=.25]{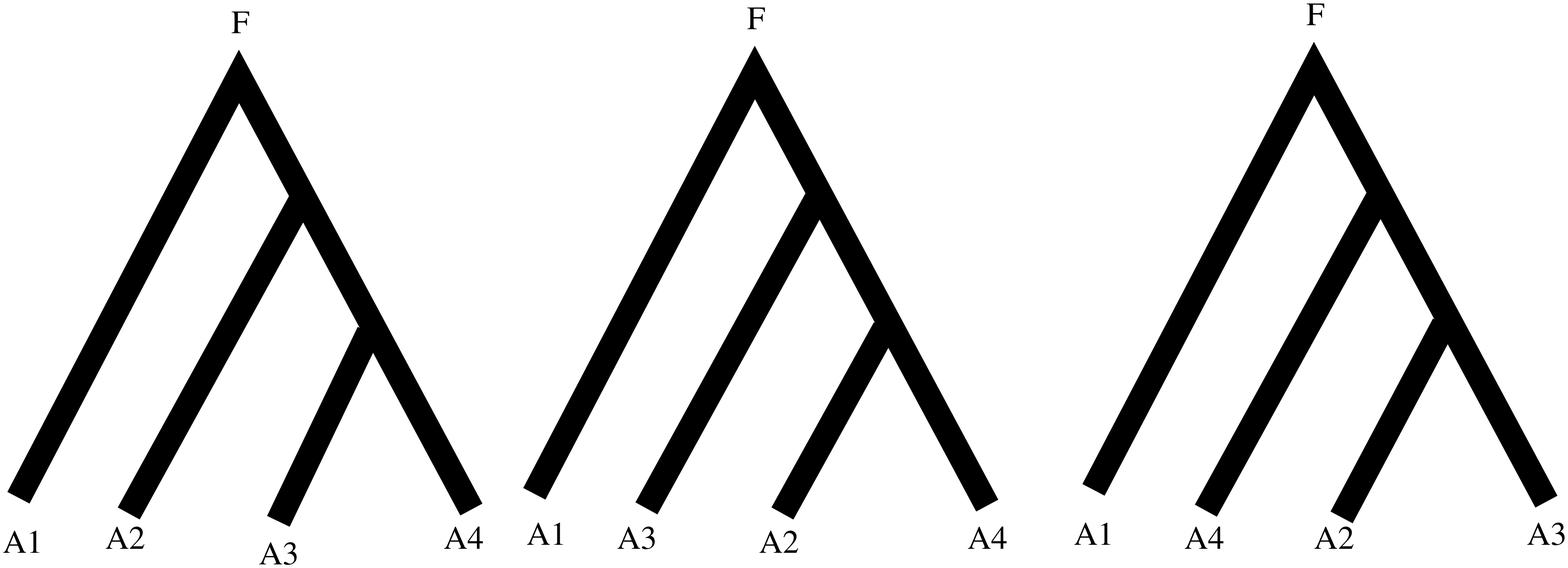}
\caption{\small{}}  
\end{center}
\end{figure}

\pagebreak

Note that there are no pictures like

\begin{figure}[!htb]\begin{center}
\includegraphics[scale=.25]{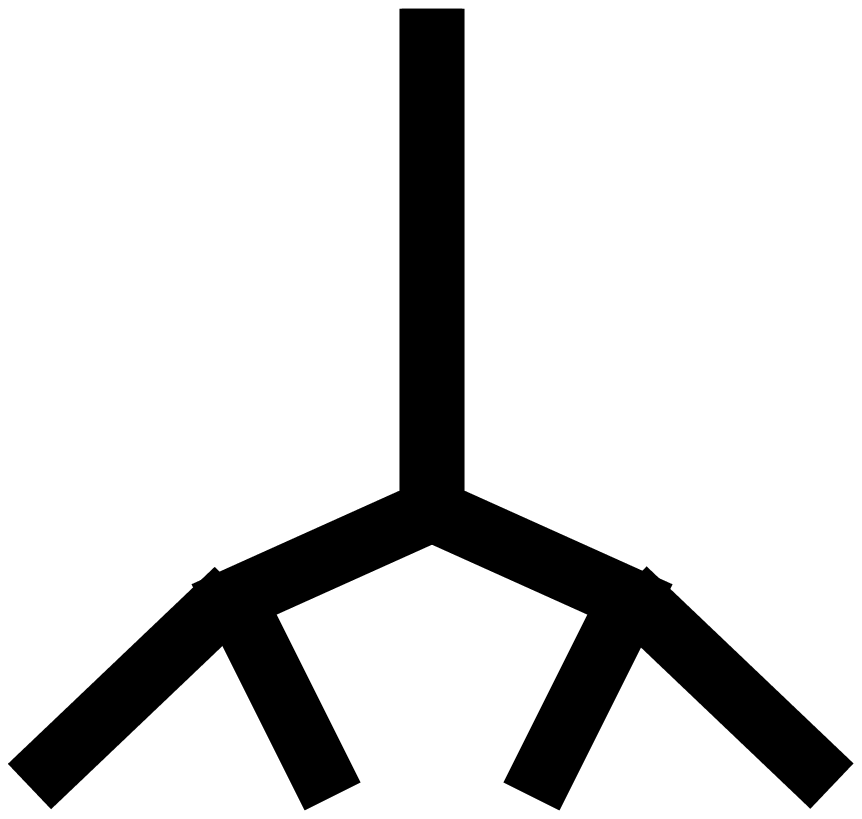}
\caption{\small{}}  
\end{center}
\end{figure}

\noindent because such give rise to equivalent algebraic statistical
models to the exhibited trees.

  We consider that
parent $F$ first gives rise to $A_1$ and $E$,
and then $E$ gives rise  to $A_2$ and $G$ and
$G$ gives rise to $A_3$  and $A_4$, as well
as the equivalent (by the discussion above) 
senarios.
The resulting algebraic statistical model
is 
$$\Sigma_{12,34}:=\s_4(Seg(\BP A_1\times \BP A_2\times \BP(A_3\ot A_4))
\cap \s_4(Seg(\BP (A_1\ot A_2)\cap \BP A_3\times \BP A_4))
$$
Similarly we get the other two possibilities
$$\Sigma_{13,24}:=\s_4(Seg(\BP A_1\times \BP A_3\times \BP(A_2\ot A_4))
 \cap \s_4(Seg(\BP (A_1\ot A_3)\cap \BP A_2\times \BP A_4))
$$
and
$$\Sigma_{14,23}:=\s_4(Seg(\BP A_1\times \BP A_4\times \BP(A_2\ot A_3))
 \cap \s_4(Seg(\BP (A_1\ot A_4)\cap \BP A_2\times \BP A_3))
$$
Note that these three are isomorphic as projective varieties,
but are situated differently in $\BP (A_1\ot A_2\ot A_3\ot A_4)$,
thus having defining equations for them would enable one
to test between different evolutionary possibilities.
An essential result
of \cite{AR2} is: 

\smallskip

{\it Once one has defining equations for
$\s_4(Seg(\pp 3\times \pp 3\times \pp 3))$, one has
defining equations for all algebraic statistical models
corresponding to bifurcating phylogenetic trees.
}

\smallskip

The proof   relies on two results.
First,   no matter how many species one
observes, because of the structure of the evolutionary
trees, the resulting algebraic statistical model
is an intersection of fourth secant varieties of
Segre varieties corresponding to summing over the four
outcomes on a hidden variable. The second (\cite{AR2}, Theorem 11)
is equivalent to (and arrived at independently of)  Proposition
\ref{lmsecbprop} below, which in particular   reduces the study
of the fourth secant variety of any triple
Segre product to the study of 
$\s_4(Seg(\pp 3\times \pp 3\times \pp 3))$.

\subsection{Entanglement and quantum computing}
In {\it quantum computing}
 (see, e.g.,  \cite{MR2230995} and the numerous references therein)
a {\it pure state} corresponds to a point of $\BP(\BC^2\otc \BC^2)$
where there are $N$ copies of $\BC^2$.
A {\it product state} corresponds
to a point of $Seg(\pp 1\times\cdots \times \pp 1)\subset\BP(\BC^2\otc \BC^2)$.
A pure state is {\it entangled} if it is not a product state, and
quantum computing is based on
exploiting entangled states. 
A perhaps overly optimistic program is to classify the
$U(2)\ctimes U(2)$ and/or $SL(2,\BC)\ctimes SL(2,\BC)$
orbits in $\BC^2\otc \BC^2$, which would 
give a complete classification of entangled states.
Failing that, one is interested in finding specific
measures of entanglement.
One measure of entanglement is called the
{\it Schmidt measure}, introduced in \cite{schmintro}. 
In the language of this paper, the Schmidt measure of a tensor is the base two log
of its rank.  
In \cite{eisertgross} they observe that
a tensor of a given Schmidt measure might be a limit
of tensors of a lower Schmidt measure, in fact they
give the explicit example of 
\eqref{Ttanten}  in their equation (19), where
their $|1,0,0>$ corresponds to $a_1\ot b_1\ot c_1$
in \eqref{Ttanten}. In \cite{eisertgross} they decompose
$ \BC^2\ot \BC^2\ot \BC^2\backslash 0$ into the union of
four disjoint components which they
label $S,B,W,GHZ$. In the language of this
paper, the components are
\begin{align*}
S&=\hat Seg(\pp 2\times\pp 2\times \pp 2)\backslash 0=
\hat Seg(\BP A\times \BP B\times \BP C)\backslash 0,\\
B&=\{\hat Seg(\BP A\times \BP (B\ot C)
\cup \hat Seg (\BP (A\ot B)\times \BP C)
\cup \hat Seg(\BP(A\ot C)\times \BP B)\}\backslash \{ 0\cup S\},
\\
W&=\hat\t(Seg(\BP A\times \BP B\times\BP C))\backslash \{0\cup S\cup B\}\\
GHZ&=\BC^2\ot \BC^2\ot \BC^2\backslash \{ 0\cup S\cup B\cup W\}.
\end{align*}
Compare $B$ with
the discussion of {\it flattenings} in \S\ref{auxvarsect}.

There is a vast literature regarding entanglement and
there does not appear yet to be a consensus regarding
what is the best way to measure entanglement, but it is clear
that secant varieties of Segre varieties
and related auxiliary varieties  are relevant for the
problem.

\section{Strassen's equations and lower bounds for
rank and border rank}\label{strasseneqnsect}

In this section we introduce Strassen's equations
and use them to give a new proof of Bl\"aser's $\frac 52$-theorem.
In \S\ref{invarstrassen} we rephrase the equations invariantly
and give generalizations.

\subsection{Strassen's equations}

Recall the notation  
  $\aaa=\tdim A$, $\bbb=\tdim B$, $\ccc=\tdim C$.

\begin{theorem}[Strassen \cite{strassen505}]\label{strasseneqn} Let $3\leq \aaa\leq \bbb=\ccc\leq r$. Let
$T\in \s_r(Seg(\BP A\times \BP B\times \BP C))$ and $\a\in A^*$ be such that
$T_{\a}:=T(\a)\in B\ot C$, considered as a map $T_{\a}:C^*\ra B$,
is of full rank. For each $\a^1,\a^2\in A^*$,
define the linear map  $T_{\a,\a^j}: B\ra B$ by  $T_{\a,\a^j}= T_{\a^j}T_{\a}\inv$. 
 Then
$$
\trank [T_{\a,\a^1},T_{\a,\a^2}]\leq 2(r-\bbb) 
$$
where   $[S,T]=ST-TS$ is the commutator of endomorphisms.
\end{theorem}

\begin{corollary}[Strassen \cite{strassen505}] $\s_4(Seg(\pp 2\times \pp 2\times \pp 2))\neq \BP (\BC^3\ot \BC^3\ot \BC^3)$.
\end{corollary}
\begin{proof}[Proof of corollary]
For generic  $T\in A\ot B\ot C=\BC^3\ot \BC^3\ot \BC^3$
and $\a,\a^1,\a^2\in A^*$, one has
  $\trank(  [T_{\a,\a^1},T_{\a,\a^2}])=3$ but for points
in $\s_4(Seg(\BP A\times \BP B\times \BP C))$, the rank is at most two.
\end{proof}

Note that an easy calculation 
with Terracini's lemma (\ref{Terracinilemma})  
  shows that $\s_4(Seg(\pp 2\times \pp 2\times \pp 2))$
is at least a hypersurface, so the above corollary shows it
is exactly a hypersurface. Strassen's equations are not presented
as polynomials above. In \S\ref{invarstrassen} we describe them as polynomials
and give generalizations. 

\begin{corollary}[Strassen \cite{strassen505}]
$\tbrank (M_{m,m,m})\geq \frac {3m^2}2$.
\end{corollary}

\begin{proof}
Write out $M_{m,m,m}$ explicitly in a good basis
and takes a generic $\a\in A^*=Mat_{m\times m}$.
Then the corresponding linear map  $T_{\a}$ is
a block diagonal matrix with blocks of size $m$, each block
identical and the entries of the block arbitrary. So we have
$\trank(  [T_{\a,\a^1},T_{\a,\a^2}])=m^2$. Hence
$m^2\leq 2(r-m^2)$ and the result follows.\end{proof}

\subsection{Proof of Bl\"aser's lower bound}\label{blaserpf}  
Here is a proof of  Theorem 
\ref{Blaser52} that uses Theorem \ref{strasseneqn},
which is implicit, but hidden, in his original proof.

\begin{lemma}\label{hypersurfsubbasislem} Let
$U$ be a vector space, let $P\in S^dU^*\backslash 0$.
Let $u_1\hd u_n$ be
a basis of $U$. Then there exists a subset $u_{i_1}\hd u_{i_s}$
of cardinality $s\leq d$ such that $P\mid_{\langle u_{i_1}\hd u_{i_s}\rangle}$
is not identically zero.
\end{lemma}

The proof is an easy exercise.

\begin{lemma}\label{Blaser52lemma} Given any basis of $Mat_{m\times m}^*$, there
exists a subset of at least $m^2-3m$ basis vectors 
that annhilate elements $Id,x,y\in Mat_{m\times m}$
such that $[x,y]:=xy-yx$ has maximal rank $m$.
\end{lemma}

\begin{proof} Let $A=Mat_{m\times m}\simeq U^*\ot W$. Fixing a  basis
  of $A^*$ is equivalent to
fixing its dual basis   of $A$.
By Lemma \ref{hypersurfsubbasislem} with $P={\rm det}$, we may
find a subset $S_1$ of at most $m$   elements
of our basis of $A$
   with some $z\in Span(S_1)$ with $\tdet(z)\neq 0$.
We use $z: U\ra W$ to identify $U\simeq W$  which enables
us to now consider $A$ as an algebra with $z$ playing
the role of the identity element.

 Now let $a\in A$ be generic. Then the map
$ad(a): A\ra A$, $x\mapsto [a,x]$ will have a one-dimensional
kernel. By letting $P=ad(a)^*(det)$ and applying 
Lemma \ref{hypersurfsubbasislem}
again, we may find a subset $S_2$ of our basis of cardinality
at most $m$ such that there is an element $x\in A$ such
that $ad(a)(x)$ is invertible. Note that $ad(x): A\ra A$ also is such
that there are elements $y$ with $ad(x)y$ invertible. Thus we
may apply Lemma \ref{hypersurfsubbasislem} a third time to find
a cardinality at most  $ m$ subset $S_3$ of our basis such
that $ad(x)y$ is invertible. Now in the worst possible case
our three subsets are of maximal  cardinality and do not
intersect, in which case we have a cardinality
$m^2-3m$ subset of our dual basis that annihilates $z=Id,x,y$ with
$\trank([x,y])=m$.
\end{proof}

\begin{proof}[Proof of Theorem \ref{Blaser52}]
Let $\phi$ denote a computation of $M=M_{m,m,m}$ of length $r$.
  Since
$\tlker(M)=0$ (i.e., $\forall a\in A\backslash 0$, $\exists b\in B$ such that
$M(a,b)\neq 0$) we may  write $\phi=\psi_1+\psi_2$
with $\trankc(\psi_1)=m^2$, $\trankc(\psi_2)= r-m^2 $
and $\tlker (\psi_1)=0$.
Now consider the $m^2$ elements of $A^*$ appearing in
$\psi_1$. Since they span $A^*$, by Lemma \ref{Blaser52lemma}
we may choose a subset of $m^2-3m$ of them that annhilate
$Id,x$ and $y$, where $x,y$ are such that $[x,y]$ has full
rank. Let $\phi_1$ denote the sum of all monomials in $\psi_1$
whose $A^*$ terms
annhilate $Id,x,y$, so $\trankc (\phi_1)\geq m^2-3m$.
Let $\phi_2=\psi_1-\phi_1+\psi_2$.

  Now apply Theorem \ref{strasseneqn} with
$T=\phi_2$, $\a=Id$, $\a_1=x$, $\a_2=y$ to get
$\tbrank(\phi_2)\geq \frac 12{\rm{rank}} [x,y]+m^2= \frac 32 m^2$
and thus $\trankc(\phi_1+\phi_2)\geq \frac 52 m^2-3m$.
\end{proof}

\section{Limits of secant planes}\label{seclimitsect}

There are several reasons for studying points on
$\s_r(Seg(\BP A_1\ctimes \BP A_n))$ that are not
on   secant $\pp{r-1}$'s. First, in order 
to prove a set of equations $E$ is a set of
 defining equations
for $\s_r(Seg(\BP A_1\ctimes \BP A_n))$, one must
prove that any point in the zero set of $E$ is either
a point on a    secant $\pp{r-1}$ or on
a limit $\pp{r-1}$. For example, the
proof of the set-theoretic GSS conjecture
  (see \S\ref{auxvarsect}) in \cite{LMsec}
proceeded in this fashion. Second,
  to prove lower bounds
for the border rank of a given tensor,
e.g., matrix multiplication, one could
try to prove first it cannot lie on
any   secant $\pp{r-1}$ and then
that it cannot lie on any limiting
$\pp{r-1}$ either. This was the technique
of proving $\tbrank(M_{2,2,2})=7$ in
\cite{Lmatrix}. Finally, 
  a central ingredient for writing 
explicit approximate algorithms is
to exploit certain   limiting
$\pp{r-1}$'s discussed below. 

This section and the next are not used in the remainder
of the article so they can be skipped by readers primarily
interested in the equations of secant varieties of Segre
varieties.
 
\subsection{Limits for arbitrary projective varieties} Let $X\subset \BP V$ be a projective variety.
Let $\s_r^0(X)$ denote the set of   points on $\s_r(X)$
that lie on a    secant $\pp{r-1}$.  
We work inductively, so we assume we know
the nature of points on $\s_{r-1}(X)$
and study points on $\s_r(X)\backslash (\s_r^0(X)\cup \s_{r-1}(X))$.

It is convenient to study the limiting $r$-planes as
  points on the cone over the Grassmannian 
in its Plucker embedding, $G(r,V)\subset \BP(\La r V)$
(see the end of \S\ref{waring}).
I.e., we consider the curve of $r$ planes as being
represented by $x_1(t)\ww \cdots\ww x_r(t)$ and
examine the limiting plane as $t\ra 0$. (There must
be a unique such plane as the Grassmannian is compact.)
 
Let $[p]\in \s_r(X)$. Then
there exist curves $x_1(t)\hd x_r(t)\subset \hat X$ with
$p\in \tlim_{t\ra 0}\langle x_1(t)\hd x_r(t)\rangle $.
We are interested in the case when
$\tdim \langle x_1(0)\hd x_r(0)\rangle <r$. 
(Here $\langle v_1\hd v_k\rangle$ denotes  the linear span
of the vectors $v_1\hd v_k$.) 
Use the notation
$x_j=x_j(0)$.    Assume 
for the moment 
that $x_1\hd x_{r-1}$ are linearly independent.
Then we may write $x_r=c_1x_1+\cdots + c_{r-1}x_{r-1}$ for some
constants $c_1\hd c_{r-1}$. Write each curve
$x_j(t)=x_j+tx_j'+t^2x_j'' +\cdots$ where derivatives are taken at $t=0$.

Consider the Taylor series  
\begin{align*}
x_1(t)\ww\cdots\ww x_r(t) = 
&(x_1+tx_1'+t^2x_1'' +\cdots)\ww \cdots \ww (x_{r-1}+tx_{r-1}'+t^2x_{r-1}'' +\cdots)
\ww (x_{r}+tx_r'+t^2x_r'' +\cdots) 
\\
%&=
%(x_1+tx_1'+t^2x_1'' +\cdots)\ww \cdots \ww (x_{r-1}+tx_{r-1}'+t^2x_{r-1}'' +\cdots)
%\ww (c_1x_1+\cdots +c_{r-1}x_{r-1}+tx_r'+\cdots ) 
%\\
%&=t(x_1'\ww x_2\ww\cdots \ww x_{r-1}\ww (c_1x_1)
%+(x_1\ww x_2'\ww \cdots \ww x_{r-1}\ww (c_2x_2)+\cdots) + t^2(...) +\cdots
%\\
&=t((-1)^r(c_1x_1' +\cdots c_{r-1}x_{r-1}'-x_r')\ww x_1\ww\cdots\ww x_{r-1}) + t^2(...) +\cdots
\end{align*}
If the $t$ coefficient is nonzero, then $p$
lies in the the $r$ plane   $\langle x_1\hd x_{r-1},(c_1x_1' +\cdots c_{r-1}x_{r-1}'-x_r')\rangle$.

If the $t$ coefficient is zero, then $c_1x_1'+\cdots + c_{r-1}x_{r-1}'-x_r'=
e_1x_1+\cdots e_{r-1}x_{r-1}$ for some constants $e_1\hd e_{r-1}$.
In this case we must examine the $t^2$ coefficient of the expansion. It is
$$ 
(\sum_{k=1}^{r-1} e_kx_k' + \sum_{j=1}^{r-1}c_jx_j''-x_r'')\ww x_1\ww\cdots\ww x_{r-1}
$$
One continues to higher order terms if this is zero.

The algorithm  of Example \ref{biniex} below uses the $t$ coefficient,
the algorithm of Example \ref{schex} uses the $t^2$ coefficient, and    Algorithm  8.2 in
\cite{schon460} uses the coefficient of  $t^{20}$!

\subsection{Limits for Segre varieties}\label{segrelimitsect}
A general curve on $\hat Seg(\BP A\times \BP B\times \BP C)$
is of the form $x(t)=a(t)\ot b(t) \ot c(t)$ where
$a(t),b(t),c(t)$ are respectively arbitrary curves
 in $A\backslash 0, B\backslash 0, C\backslash 0$
with $a(0)=a$ etc.
We have $x'=a'\ot b\ot c + a\ot b'\ot c + a\ot b\ot c'$
where $a',b',c'$ are respectively arbitrary
elements of $A,B,C$, and higher order derivatives
are obtained similarly.

While the easiest way to obtain $r$ points
that  are linearly  dependent in the limit
is to have two points limit to the same point, this
turns out to be not as useful for  upper bound
algorithms as more
subtle limits. On the other hand,  when $r$ is sufficiently small,
any other
type of limit
involves exploiting the   geometry of the Segre variety as
we now explain.

To simplify the situation, we  
work inductively  and just look
at \lq\lq primitive\rq\rq\ cases, i.e., require
  that the points on the limiting
$\pp{r-1}$ do not lie on
$\s_r(Seg(\BP A'\times \BP B'\times \BP C'))$
where $\tdim A'\leq \tdim A$ etc... (with at
least one inequality strict), and moreover
that the points do not lie on
$\s_{r-1}(Seg(\BP A\times \BP B\times \BP C))$.

For example,   for the two factor
Segre $Seg(\BP A\times \BP B)$, (which, if we are working by induction,
must be studied for the three factor case, as it
corresponds to the case $\tdim C'=1$),   in order to have
$x_1\hd x_r\in Seg(\BP A\times \BP B )$
such that
$\tdim\langle x_1\hd x_r\rangle<r-1$
and 
 the points are not contained in some
$Seg(\BP A'\times \BP B' )$,
we must have $\aaa+\bbb\leq r$ (see the erratum
to \cite{Lmatrix}).
In the erratum to \cite{Lmatrix} we determine
all possible $x_1\hd x_6\in Seg(\pp 3\times\pp 3\times
\pp 3)$ with $\tdim \langle x_1\hd x_6\rangle <6$.
The only possible cases where the points
fail to lie in some $Seg(\pp 0\times \BP B\times \BP C)$
occur when they all lie in some $Seg(\pp 2\times \pp 2\times\pp 2)$.

A basic property of projective space
is that if $X^n\subset\pp\na$ is a subvariety,
then a general $\pp{a}$ will intersect $X$
in $\tdeg(X)$ points. (In fact this
is the definition of the degree of $X$.)
One can calculate that $\tdeg(Seg(\pp 2\times \pp 2))=6$
(see, e.g., \cite{Harris}, lecture 18) and
$\tcodim (Seg(\pp 2\times \pp 2))=4$.
Therefore, for any set of $5$ points on
$Seg(\pp 2\times \pp 2)$ that are linearly independent,
i.e., that span a $\pp 4$,   there
is a sixth point in the $\pp 4$ that also lies
on the Segre. Taking the span of these six
points as our $x_i(0)$,  we get a   limit set
that allows the use of derivatives.
This type of limit set is used
several times in Example \ref{schex} to build
Sch\"onhage's approximate algorithm for
multiplying $3\times 3$ matrices using $21$ multiplications.

Simliarly $\tdeg(Seg(\pp 1\times\pp 1\times \pp 1))= 6$  
and $\tcodim(Seg(\pp 1\times\pp 1\times \pp 1))=4$, 
which is exploited  in Example \ref{biniex}.

\section{Upper bounds}\label{upperbndsect}

We now discuss how to use  the geometry
discussed above to find explicit approximate
algorithms for executing a bilinear map.

\subsection{Sch\"onhage's results}\label{Schonsect}  
Sch\"onhage \cite{schon460}  isolated a common aspect to certain
approximate algorithms for matrix multiplication which
enabled him to generalize them and prove  
upper bounds for the exponent of matrix multiplication
without even having explicit 
approximate algorithms. The essence of
his idea is as
follows: 

\smallskip

{\it Say we have two bilinear maps   $f:U^*\times V^*\ra W$ and $g: \tilde U^*\times \tilde V^*\ra \tilde W$.
Under certain
conditions,   $\tbrank(f\oplus g)<\tbrank (f) +\tbrank (g)$,
where $f\oplus g: (U\op \tilde U)^*\times (V\oplus\tilde V)^*
\ra(W\oplus \tilde W)$.
}

\smallskip

Letting $A=U\op \tilde U $, $B=V\op \tilde V$, $C=W\op \tilde W$, recall that
  curves $x_j(t)$ on $\hat Seg(\BP A\times \BP B\times \BP C)$ are of the form $a_j(t)\ot b_j(t)\ot c_j(t)$.
We will   obtain an approximate algorithm for
$f\oplus g$ by    having the $a_j(0)$ be the $U$ vectors needed for 
the $f$ factor,
the $a_j(0)'=0$, and the $a_j(0)''$ be the $\tilde U$ vectors needed for the $g$ factor.
Then for the $B$ factor we take  the
$b_j(0)$ to be the $V$ vectors needed for $f$ and the $b_j(0)'$ the
$\tilde V$ vectors needed for $g$, and the $C$ limits are of the same nature as
the $B$ limits. Then the sum of the second derivatives will be $f\op g$. The only
problem is,
as explained in \S\ref{seclimitsect}, we need   the zero-th and first order terms to be
linearly dependent so that we are allowed to take the sum of the
second derivatives. To obtain linear dependence, the points   must lie in some degenerate position
with respect to the Segre, but   this is difficult to arrange.
Sch\"onhage's solution is to have these limit points in a two factor Segre (where
it is easier to have degenerate limits), but this forces one of
each $U,V,W$ and $\tilde U,\tilde V,\tilde W$ to be one-dimensional. 
Moreover, these restrictions only
take  care of the zero-th order term.   
To get the first order term killed,   two 
of e.g.,  $\tilde U,\tilde V,\tilde W$
are taken to be of dimension one and the third, say $\tilde W$ to
be of dimension roughly $\tdim U\tdim V$ (assuming $\tdim W=1$).
Even so, we still must add in a few extra terms
to insure linear dependence, but they are small  in number.
Sch\"onhage points out that in this situation
it is known that  neither of the 
$f,g$   admits
an approximate algorithm better than the standard algorithm.
A more geometric understanding of
this \lq\lq trick\rq\rq\ could lead to better upper bounds.
What follows are two examples for matrix multiplication, the second
of which follows   the above scheme.

\begin{example}[Bini et. al.]\label{biniex} An approximate algorithm for multiplying
$2\times 2$ matrices where the first matrix has a zero in the
$(2,2)$ slot is presented in \cite{MR534068}. In
what follows we show how the algorithm   corresponds to a
point of $\s_5(\pp 2\times \pp 3\times \pp 3)$. (It is
relatively simple to pass back and forth between the algorithms
and the description of the 
limiting $\pp{4}$ that lies in $\s_5(Seg(\pp 2\times \pp 3\times \pp 3))$ that the tensor lies on. But
the description of the $\pp 4$   shows the
non-uniqueness of the algorithm and the salient geometric
facts that are used more transparently.)
In this case we have $5$ points that are linearly dependent.
In fact only four are needed, one can take any $5$-th point in
the span of the four and ignore it as its derivatives are not
needed for the algorithm. 
We take
$$
x_1=\aa 12\ot \bb 12\ot \cc 12,\ x_2=\aa 21\ot \bb 11\ot \cc 11,\
x_3=\aa 12 \ot \bb 12\ot (\cc 11+\cc 12),\
 x_4=\aa 21\ot (\bb 11+\bb 12)\ot \cc 11.
$$
Note that all these points lie on a $Seg(\pp 1\times \pp 1\times \pp 1)$.
Because $\tcodim(Seg(\pp 1\times \pp 1\times \pp 1))=4$, we are assured there is a fifth point  of $Seg(\pp 1\times  \pp 1\times \pp 1)$ in the
span of these four. (A general $\pp 3$ will
intersect $Seg(\pp 1\times \pp 1\times \pp 1))$ in
$\tdeg(Seg(\pp 1\times \pp 1\times \pp 1))=6$ points.) 
Moreoever,  the $5$-th point will not be
in the span of any three of $x_1\hd x_4$.
Then taking
\begin{align*}
x_1'&=\aa 11\ot \bb 12\ot \cc 12+\aa 12\ot \bb 22\ot \cc 12-\aa 12\ot \bb 21\ot \cc 12,
\ \
x_2'=\aa 11\ot \bb 11\ot \cc 11+\aa 21\ot \bb11\ot \cc21-\aa 21\ot \bb 11\ot \cc 22,
\\
x_3'&=\aa 12\ot \bb 21\ot (\cc 11+\cc 12),\
x_4'=\aa 21\ot (\bb 11+\bb 12)\ot \cc 22,
\end{align*}
  our matrix multiplication operator $M$
for the partially filled matrices is $M=x_1'+x_2'+x_3'+x_4'$.  The fact that
we didn't use any of the initial points is not suprising as
the derivatives can always be altered to incorporate the
initial points.

\smallskip

A splitting of the computation is the key to
the reduction here as well.  Split the calculation
of $M$ into two pieces, the terms involving
$\aa 11$ and the rest. Those terms  involving $\aa 11$ can
be accomplished using two multiplications and
the rest can be accomplished using six.
We change notation slightly and write
$x_j=a_j\ot b_j\ot c_j$ and $x_j'=a_j'\ot b_j\ot c_j + a_j\ot b_j'\ot c_j+a_j\ot b_j\ot c_j'$
as we did before we began this example.
The elements of $B\ot C$ appearing with $\aa 11$
each appears in the original $x_1,x_2$, so in order
to have them appear in the final tensor we just
need to take $a_1',a_2'=\aa 11$.
Now to have the terms involving $\aa 12,\aa 21$ appear
in the final tensor, we need to differentiate the
  terms on the $B$ and $C$ factors.
We can obtain two of these by setting $b_1'=\bb 22$
and $c_2'=\cc 21$. We can get the remaining terms
using $x_3'$ and $x_4'$ but we must introduce
an error, which can then be absorbed by modifying
$b_1'$ and $c_2'$. The result is that
$a_1'=a_2'=\aa 11$, $b_1'=\bb 22-\bb 21$,
$c_2'=\cc 21-\cc 22$, $b_3'=\bb 21$, $c_4'=\cc 22$
and all the other first derivatives are zero.
\end{example}

\begin{remark} There is a similarity between this example and
the   algorithms   using multiplicative
complexity discussed in \S\ref{hopkerex}.
\end{remark}

\begin{example}[Sch\"onhage]\label{schex}
Consider matrix multiplication of $3\times 3$ matrices
where in the first matrix   $\aa 21=\aa 31=0$,
in the second that $\bb 22=\bb 23=\bb 32=\bb 33=0$, and thus
  $\cc 22=\cc 23=\cc 32=\cc 33=0$ as well.
We again split the computation into terms involving
$\aa 11$ and those that do not. (It might be  
  useful to think of this multiplication
as $B\times C\ra A$ to make it look more symmetric.)
Those that do not involve $\aa 11$ use $6$ multiplications
in the na\"\i ve algorithm and those involving $\aa 11$ use
four. 

As explained in \S\ref{segrelimitsect},   $\pp 4\cap (Seg(\pp 2\times \pp 2))$
will generally consist of   $6=\tdeg(Seg(\pp 2\times \pp 2))$
points.
Now the   principle described above is used. That is, the initial
$6$ terms contain the correct six monomials in the $B,C$ factors
for the terms without $\aa 11$ and the second derivatives of the
$A$ factor in these terms are used to provide the correct $A$ terms, while
the original $A$ factor term is always $\aa 11$ and
it is paired with the derivatives in the $B,C$ factors of the
original terms.
In this example,
the spaces in $B,C$ where the two different pieces live
are nearly disjoint, so we need to differentiate twice
to be able to get both the $B$ and $C$ coefficients new
(which is why we used second, rather than first derivaties
in the $A$-factor).

What is interesting about this example is that
taking three such blockings, one can \lq\lq cover\rq\rq\
the space of three by three matrices, and adding them
together    obtain  an approximate algorithm for
$M_{3,3,3}$ using  $21$ multiplications.
\end{example}

\subsection{Finite group   approach to upper bounds}\label{finitegpsect}
Cohn and Umans \cite{CU} have proposed a   different approach to constructing
  algorithms for matrix multiplication using the discrete
Fourier transform and the representation theory of finite groups.

Let $G$ be a finite group and $\BC [G]$ its group algebra.
(See e.g., \cite{Serre} for   definitions and properties
of the group algebra.) The
discrete Fourier transform (DFT) $D: \BC [G]\ra \BC^{|G|}$ is
an invertible linear map that actualizes 
Wedderburn's theorem  that $\BC [G]\simeq Mat_{d_1\times d_1}(\BC)
\times\cdots\times Mat_{d_k\times d_k}(\BC)$, where
$G$ has $k$ irreducible representations and the dimension (character)
of the $j$-th is $d_j$.
(See e.g., \cite{BCS} for an exposition.)
Thus multiplication in the group ring is reduced
to 
  multiplication
 of $d_1\times d_1\hd d_r\times d_r$ matrices.   

The idea is, to multiply $Mat_{n\times m}\times Mat_{m\times p}\ra
Mat_{n\times p}$ one first   bijectively maps bases of each of these three
spaces into  subsets of some finite group $G$. The subsets 
are themselves formed from three subsets $S_1,S_2,S_3$, of cardinalities
$n,m,p$ which have a disjointness property,
 called the {\it triple product property} in \cite{CU}:
if $s_1s_2s_3=Id$, with $s_i\in S_i\inv S_i$, then each $s_i=Id$.
Then the maps are to 
the three subsets $S_1\inv S_2$, $S_2\inv S_3$, $S_1\inv S_3$.
The triple product property enables one to read off matrix multiplication
from multiplication in the group ring.
They then show, if $\o$ is the exponent of matrix multiplication,
 that, if one can find such a group and subsets, then
$$
(nmp)^{\frac \o 3}\leq d^{\o -2}|G|
$$
where $d$ is the largest character of $G$.
So one needs to find groups that are big enough to support
triples satisfying the triple product property but as small
as possible and with largest character as small as possible.

In \cite{CKSU} they give explicit examples which recover
$\o <2.41$ and state several combinatorial and group theoretic
conjectures  that,  if true, would imply $\o=2$.

\section{Dimensions of secant varieties of Segre varieties}\label{secdimsect}

The most basic invariant of
an algebraic variety is its dimension. In this
section we discuss the standard tool for computing
dimensions of secant varieties of projective varieties
and its application to secant varieties of Segre varieties.
The results of this section are not used in the following sections.
 
\subsection{Dimensions of secant varieties of Segre varieties
and matrix multiplication}
Let $A,B,C$ be vector spaces of dimensions $\aaa,\bbb,\ccc$.
By Remark \ref{secdimexpect},
 the expected
dimension of $\s_r(Seg(\BP A\times \BP B\times \BP C))$
is $r(\aaa-1+\bbb-1+\ccc-1)+r-1= r(\aaa+\bbb+\ccc-2)-1$. The dimension
of the ambient space is $\aaa\bbb \ccc-1$, so we expect
$\s_r(Seg(\BP A\times \BP B\times \BP C))$ to fill $\BP (A\ot B\ot C)$ as soon as
$r(\aaa+\bbb+\ccc-2)-1\geq \aaa\bbb \ccc-1$, i.e.,
\begin{equation}\label{filldim}
r\geq \frac{\aaa\bbb\ccc}{\aaa+\bbb+\ccc-2}.
\end{equation}

Note that in the case $\aaa=\bbb=\ccc$ equation
\eqref{filldim} becomes $r\geq \aaa^3/(3\aaa-2)\simeq \aaa^2/3$. Taking $\aaa=n^2$, the right hand side of \eqref{filldim} is
roughly $n^4/3$, showing already that matrix multiplication
is far from being a generic bilinear map, as even the standard
algorithm gives $\trankc(M_{n,n,n})\leq n^3$. (The actual typical
$X$-rank cannot be smaller than the expected typical $X$-rank.)
However for $n=2$ we obtain $r\geq 64/10$ and thus $r=7$ is expected
to (and we will see below does) fill, so $M_{2,2,2}$ is generic
in this sense.

\subsection{Terracini's lemma and applications} 
\label{secsegdimenssect}
Recall the notations from the begining of \S\ref{secantvarsect} and
adopt the additional notation that
for $Z\subset \BP V$,  $\hat T_{[z]}Z = T_z\hat Z\subset V$ is the embedded
tangent space to $\hat Z$ at $z\in\hat Z$.

\smallskip

\begin{lemma}[Terracini's Lemma (see, e.g., \cite{ciro,IvL,zak}) ]\label{Terracinilemma} 
If $[x]\in J(Y, Z)\subsmooth$
with $[x]=[y+z]$, such that $[y]\in Y\subsmooth, [z]\in Z\subsmooth$,
then
$$
\hat T_{[x]} J(Y, Z) = \hat T_{[y]}Y + \hat T_{[z]} Z.
$$
Thus, if $[p]=[x_1+\cdots +x_r]\in \s_r(X)_{smooth}$ with
$[x_j]\in X_{smooth}$, then
$$
\hat T_{[p]} \s_r(X) = \hat T_{[x_1]}X +\cdots  +\hat T_{[x_r]}X.
$$
\end{lemma}

Terracini's lemma implies that for a variety
$X\subset \BP V$, if any given $\s_r(X)$ is nondegenerate
(i.e. of the expected dimension) and
of dimension $r\tdim X+r-1$,
then all   $\s_{r'}(X)$ for $r'<r$ are nondegenerate.

Thus   one can show all
secant varieties of $X$ are non-degenerate if
one shows  $\s_p(X)=\BP V$ if $\tdim \BP V=p(n-1)+p-1$.

The following trick occurs frequently in the literature: let
$Y_1\hd Y_p\subset X$, so
$\hat T_{y_1}Y_1+\cdots +\hat T_{y_p}Y_p\subseteq\hat T_{[y_1+\cdots+ y_p]}
\s_p(X)$. If one can show $\hat T_{y_1}Y_1+\cdots +\hat T_{y_p}Y_p=V$,   one
has shown $\s_p(X)=\BP V$. Lickteig and Strassen show that
for $X=Seg(\BP A\times \BP B\times \BP C)$, remarkably
just taking the $Y_i$ to be the Segre itself at most three times and
taking other the $Y_i$ to be linear spaces
in it is sufficient for certain cases:

\begin{lemma}[Lickteig \cite{lick}]\label{licklem} Adopt   the notation 
  $\BP A_i=\BP (A\ot b_i\ot c_i)\subset Seg(\BP A\times \BP B\times \BP C)$, $\BP B_j=\BP (a_j\ot B\ot c_j')\subset Seg(\BP A\times \BP B\times \BP C)$.

\begin{enumerate}
\item  \label{licklem1}
We may choose points $a_1\hd a_s\in A$, $b_1\hd b_q\in B$,
$c_1\hd c_q,c_1'\hd c_s'\in C$, 
such that
$$
\hat J(\BP A_1\hd \BP A_q,\BP B_1\hd \BP B_s)= A\ot B\ot C
$$
when   $q=\bbb l_1$, $s=\aaa l_2$ and $\ccc=l_1+l_2$
  and when 
  $\aaa=\bbb=2$,  $q+s=2\ccc$,  $s,q\geq 2$.
 
\item  \label{licklem2}
We may choose points $a_1\hd a_s\in A$, $b_1 \hd b_q\in B$,
$c_1\hd c_q,c_1'\hd c_s'\in C$, 
such that
$$
\hat J(\s_2(Seg(\BP A\times \BP B\times \BP C)), \BP A_1\hd \BP A_q,\BP B_1\hd \BP B_s)= A\ot B\ot C
$$
when  $q+s+2=\ccc$ and $\aaa=\bbb=2$.
 
\item
We may choose points $a_1\hd a_s\in A$, $b_1 \hd b_q\in B$,\\
$c_1\hd c_q,c_1'\hd c_s'\in C$,  
such that
$$
\hat J(\s_3(Seg(\BP A\times \BP B\times \BP C)), \BP A_1\hd \BP A_q,\BP B_1\hd \BP B_s)= A\ot B\ot C
$$
when  $q=s=\ccc-2\geq 2$ and $\aaa=\bbb=3$.
\end{enumerate}
\end{lemma}

Using Lemma \ref{licklem}, Lickteig shows 

\begin{theorem}[Lickteig \cite{lick}] $\s_r(Seg(\BP A\times \BP B\times \BP C))$ is
nondegenerate for all $r$ whenever $\aaa\leq \bbb\leq \ccc$, 
$\bbb,\ccc$ are even and $\aaa\bbb\ccc/(\aaa+\bbb+\ccc-2)$ is an integer.
\end{theorem}

With a little more work one obtains
Theorem \ref{licksecdims}.

\medskip

A classical technique for showing a secant variety
of any variety $X\subset\BP V$
is degenerate is to find a variety
$Y\subset \BP V$, with $X\subset Y$, with $\s_k(Y)$ very
degenerate. Then, if $X$ \lq\lq catches up\rq\rq\ 
i.e., if there exists $r$ such that $\s_r(X)=\s_r(Y)$, then
$\s_t(X)=\s_t(Y)$ for all $t>r$ as well. (See, e.g.
\cite{CGGpreprint} for a recent application.)  To see this,
first note that
for $u<r$,
$\s_r(X)=J(\s_{r-u}(X),\s_u(X))\subseteq 
J(\s_{r-u}(Y),\s_u(X))\subseteq\s_r(Y)$,
so $\s_r(Y)= 
J(\s_{r-u}(Y),\s_u(X))$.
Now  write $t=mr+u$,
\begin{align*}
\s_t(X)&=J(\s_{mr}(X),\s_u(X))\\
&=J(\s_{(m-1)r}(Y),\s_{u}(Y),J(\s_{r-u}(Y),\s_u(X)))\\
&=\s_{mr+u}(Y).
\end{align*}

\smallskip

In particular, since $\s_r(Seg(\BP A\times \BP B))$ is
very degenerate, if we have a three factor
case that is \lq\lq unbalanced\rq\rq\ in the sense
that one space is much smaller than the others, it
can catch up to a corresponding two factor case.
For example $\s_2(Seg(\pp 1\times \pp 1\times \pp 3))=
\s_2(Seg(\BP(\BC^2\ot\BC^2)\times \pp 3))$.
Note that when this catching up occurs, if one
knows the ideal of the {\it a priori} larger variety,
one obtains the ideals of the secant varieties of the
smaller variety.
Other uses of 
auxiliary varieties to understand the secant varieties
of Segre varieties, are discussed in  
in \S\ref{auxvarsect}.

\smallskip

  In the past few years there have been several papers on
the dimensions of secant varieties of Segre varieties, e.g.,
\cite{MR2202248,MR1930149,MR1898833,CGGpreprint,AOP}.
These papers use  methods similar to those of Strassen
and Lickteig, but the language is more geometric 
(fat points, degeneration arguments). Some explanation
of the relation between the algebreo-geometric
and  tensor language is given in \cite{AOP}.

With such steady progress, it seems reasonable to
hope for a complete solution for the secant defectivity
of Segre varieties in the near future, at least in
the three factor case.

\section{Invariant description of Strassen's equations and generalizations}
\label{invarstrassen}

In this section we first rephrase Strassen's equations
as the image of a $GL(A)\times GL(B)\times GL(C)$-equivariant
map. We use this rephrasing to describe how to explicitly
write a basis of his equations in a \lq\lq good\rq\rq\ basis
and to generalize his equations. To ease the reader
into this perspective, we begin with a familiar case.

\subsection{Warm up: Invariant description of generators
of the ideal of $\s_r(Seg(\BP A\times \BP B))$}\label{warmup}
The set of $\aaa \times \bbb$ matrices of rank at most $r$
is the zero set of the $(r+1)\times (r+1)$ minors, in
fact these minors generate the ideal of $\s_r(Seg(\BP A\times \BP B))$.
To understand this space of equations invariantly,
we begin with two by two minors. Choose bases
$\{ a_i\}$ of $A$, $\{ b_s\} $ of $B$ and write our resulting matrix
representing a point of $A\ot B$ as
$X=(x^i_s)$. Consider
the minor $P_{ij,st}:=x^i_sx^j_t-x^i_tx^j_s\in S^2(A\ot B)^*$. Note that 
$P_{ij,st}=-P_{ji,st}$ and $P_{ij,st}=-P_{ij,ts}$.
Hence   $P_{ij,st}\in\La 2A^*\ot \La 2 B^*$, and in fact
we have an injective map
$$\La 2A^*\ot \La 2 B^*\ra S^2(A\ot B)^*$$ 
whose image is
the space of $2\times 2$ minors. By the same reasoning,
there is an injective map
$\La dA^*\ot \La dB^*\ra S^d(A\ot B)^*$ with image
the $d\times d$ minors. We conclude

\smallskip

{\it The ideal of $\s_r(Seg(\BP A\times \BP B))$ is generated
by $\La{r+1}A^*\ot \La{r+1}B^*\subset S^{r+1}(A\ot B)^*$.}

\smallskip

We will see in  \S\ref{reptheorysect}  that 
$\La{r+1}A^*\ot \La{r+1}B^*$ is an irreducible
$GL(A)\times GL(B)$-submodule of $S^{r+1}(A\ot B)^*$. 
A more precise goal than \lq\lq finding equations for
secant varieties of Segre varieties\rq\rq\  is to find
the irreducible modules generating their ideals.
When we discuss finding invariant descriptions of
sets of equations, ultimately we will mean as modules,
but in the interm, we can simply mean \lq\lq without reference
to choices of bases\rq\rq , such as we have done here
for the $(r+1)\times (r+1)$ minors.

\subsection{Strassen's equations reconsidered}\label{strassenrecon}
In order to understand Strassen's equations  
  invariantly,
we would like to get rid of the choices of $\a,\a^1,\a^2$, and  
the requirement that $\a$ is such that $T(\a)$ be invertible
in Theorem \ref{strasseneqn}. In what follows we will
deal with tensors instead of endomorphisms, composition
of endomorphisms will correspond to contractions of tensors,
and the commutator of two endomorphisms will correspond
to contracting a tensor in two different ways and
taking the difference of the two results.
Note that matrix multiplication 
$M: (U^*\ot V)\times (V^*\ot W)\ra U^*\ot W$
itself
is simply the contraction of $V$ with $V^*$,

A linear map $f: V\ra W$ induces   linear maps $f^{\ww k}:\La k V\ra \La k W$. If $\tdim V=\tdim W=n$ then,
letting $\tdet(f):=f^{\ww n}$, we have $f^{\ww n-1}=f\inv \ot \tdet(f)$,
which follows from the  canonical identification $\La{n-1}V\simeq V^*\ot \La n V$.

The punch line of this section is 

{\it  Strassen's equations
correspond to the image of the composition
  of the inclusion
$$
\La 2 A\ot S^{\bbb-1}A\ot \La \bbb B\ot B\ot \La \bbb C\ot C
\ra (A\ot B\ot C)^{\bbb+1}
$$
with the projection
$$
(A\ot B\ot C)^{\bbb+1}\ra S^{\bbb+1}(A\ot B\ot C).
$$
}

We remark that the composition of these two maps is
not   injective. In \S\ref{strassmodulesect} we describe the image
precisely. We emphasize this perspective  because
it leads to vast generalizations of Strassens equations
discussed in \S\ref{genstrsect}.

\medskip

Given $T\in A\ot B\ot C$, recall
our notation $T_{\a}\in B\ot C$.
We have $T_{\a}^{\ww \bbb-1}\in \La{\bbb-1}B\ot \La{\bbb-1}C=
\La{\bbb-1}B\ot C^*\ot \La \bbb C$.  We may wedge
the $\La{\bbb-1}B$ and $B$ factors in
$$T_{\a}^{\ww\bbb-1}\ot T_{\a^j}
\in\La{\bbb-1}B\ot C^*\ot \La \bbb C\ot   B\ot C
$$
together to obtain  
an element 
$$T^{\a}_{\a^j}\in \La \bbb B\ot C^*\ot \La \bbb C\ot C
=C^*\ot C\ot \La \bbb B\ot \La \bbb C.
$$
That is, up to tensoring with a one-dimensional
vector space, we have a linear maps $C\ra C$ and
can now take their commutators.
Consider 
$$T^{\a}_{\a^1}\ot T^{\a}_{\a^2}
\in (\La \bbb B\ot C^*\ot \La \bbb C\ot C)^{\ot 2}
=C^*\ot C\ot C^*\ot C\ot (\La \bbb B)^{\ot 2}\ot (\La \bbb C)^{\ot 2}
$$
and contract a copy of $C$ from $T^{\a}_{\a^1}$ with a copy of $C^*$
from $T^{\a}_{\a^2}$
to obtain an element of 
$C^*\ot C\ot (\La \bbb B)^{\ot 2}\ot (\La \bbb C)^{\ot 2}$. This contraction
  corresponds to the matrix multiplication
of $T^{\a}_{\a_1}$ with $T^{\a}_{\a_2}$. And  reversing
the roles of $T^{\a}_{\a^1},T^{\a}_{\a^2}$ reverses
the order  of the matrix multiplication. Thus
  the difference of these two contractions  is  
$$
[T^{\a}_{\a^1},T^{\a}_{\a^2}]\in C^*\ot C\ot  (\La \bbb B)^{\ot 2}\ot (\La \bbb C)^{\ot 2}
$$
and Strassen's theorem states that the rank of $[T^{\a}_{\a^1},T^{\a}_{\a^2}]$
is at most $2(r-\bbb)$.

\smallskip

With a little more care, one obtains
a lower degree tensor, see \cite{LMsecb} for
details.

\begin{remark} Strassen's equations were rediscovered in \cite{AR2},
guided by the geometry of phylogenetic trees, which also enabled
a nice presentation of them.
The recent preprint \cite{ottnew} gives
an even simpler  description of Strassen's equations.
Unfortunately  the generalizations discussed below are not
evident from either of these presentations.
\end{remark}

\subsection{Explicit polynomials in bases} Here are polynomials corresponding to  
  Strassen's commutator being of rank at most $w$:  Let $\a^1,\a^2,\a^3$
be a   basis of $A^*$, $\b_1\hd \b_{\bbb}$, $\xi_1\hd \xi_{\bbb}$ bases of
$B^*,C^*$. Consider the element
$$
P=\a^2\ww\a^3\ot (\a^1)^{{\bbb}-1}\ot \b_1\wcdots \b_{\bbb}
\ot \b_s\ot \xi_1\wcdots \xi_{\bbb}\ot \xi_t
$$
This expands to (ignoring scalars)
\begin{align*}
&(\a_2\ot \a_3-\a_3\ot \a_2)\ot (\a_1)^{{\bbb}-1}\ot
(\sum_j (-1)^{j+1}\b_{\hat j} \ot \b_j\ot \b_s)\ot
(\sum_k (-1)^{k+1}\xi_{\hat k} \ot \xi_k\ot \b_t)\\
&=
(-1)^{j+k}[
((\a_1)^{b-1}\ot \b_{\hat j}\ot \xi_{\hat k} )
\ot (\a_2\ot \b_j\ot \xi_t)\ot (\a_3\ot \b_s\ot \xi_k)\\
&\ \ \ -
((\a_1)^{b-1}\ot \b_{\hat j}\ot \xi_{\hat k} )
\ot (\a_3\ot \b_j\ot \xi_t)\ot (\a_2\ot \b_s\ot \xi_k)].
\end{align*}
A hat over an index indicates the wedge product of all
vectors in that index range except the hatted one.
If we choose dual bases for $A,B,C$ and write
$T=a_1\ot X + a_2\ot Y + a_3\ot Z$ where
the $a_j$ are dual to the $\a_j$ and $X,Y,Z$
are represented as
$b\times b$ matrices with respect to the dual bases of $B,C$,
then, let $P(T)$ be the matrix with 
$$
P(T)^s_t=
\sum_{j,k}(-1)^{j+k}(\tdet X^{\hat j}_{\hat k})
(Y^j_tZ^s_k-Y^s_kZ^j_t)
$$
where $X^{\hat j}_{\hat k}$ is $X$ with its $j$-th row and
$k$-th column removed.   Strassen's commutator has rank at most $w$ if and
only if 
all the $(w+1)\times(w+1)$ minors of $P(T)$ are zero.
It turns out that when one takes the determinant of
$P(T)$, one
gets a reducible polynomial that is divisible by
the determinant of $X$, so, e.g., when $b=3$ one
obtains an irreducible polynomial of degree nine
(as opposed to $12$).

 \subsection{Generalizations  of  of Strassen's conditions}
\label{genstrsect}
 The key point
in the discussion above was that contracting
$T$ in two different ways yielded tensors that  
commute  if $T$ is in $\s_r(Seg(\BP A^*\times \BP B^*\times \BP C^*)$.
Consider, for $s,t$ such that $s+t \leq \bbb$ and $\a,\a_j\in A^*$,
the tensors
$$T_{\a_j}^{\ww s}\in \La sB\ot \La s C,\ 
T_{\a}^{\ww t}\in \La tB\ot \La t C
$$
(in \S\ref{strassenrecon} we had  $s=1, t=\bbb-1$).
We   contract $T_{\a}^{\ww t}\ot T_{\a_1}^{\ww s}\ot T_{\a_2}^{\ww s}$ to
obtain   elements of $\La {s+t}B\ot \La {s+t}C\ot \La s  B\ot \La s C$ in two
different ways, call these contractions $\psi^{s,t}_{\a,\a_1,\a_2}(T)$ and
$\psi^{s,t}_{\a,\a_2,\a_1}(T)$. 

Now say $\trankc (T)=r$ so we may write $T=a_1\ot b_1\ot c_1+\cdots + a_r\ot b_r\ot c_r$
for elements $a_i\in A$, $b_i\in B$, $c_i\in C$. We have
$$\psi^{s,t}_{\a,\a_1,\a_2}(T)=\sum_{|I|=s,|J|=t,|K|=s}
\langle a_{I},\a_1\rangle\langle a_{J},\a\rangle \langle
a_{K},\a_2\rangle (b_{I+J}\ot b_{K})\ot (c_{I}\ot
c_{J+K}),$$ 
where 
$a_I=a_{i_1}\ww\cdots\ww a_{i_s}\in\La sA$,
$\langle A_I,\a\rangle\in \La{s-1}A$
and   $a_{I+J}=a_I\wedge a_J$ etc.  For
this to be nonzero, we need $I$ and $J$ to be disjoint subsets of
$\{1,\ldots ,r\}$. Similarly, $J$ and $K$ must be disjoint.
If $s+t=r$ this implies $J=K$.
In summary:

\begin{theorem}\cite{LMsecb} For $T\in \s_{s+t}(Seg(\BP A\times \BP B\times\BP C))$, 
for all $\a,\a_1,\a_2\in A^*$
$$
\psi^{s,t}_{\a,\a^1,\a^2}(T)-\psi^{s,t}_{\a,\a^2,\a^1}(T)=0.
$$ 
\end{theorem}

We  have the bilinear map
$$
(\La 2(S^sA)\ot S^tA)^*\times (A\ot B\ot C)^{\ot 2s+t}\ra \La{s+t}B\ot \La{s+t}C
\ot \La sB\ot \La sC.
$$
whose image is $\psi^{s,t}_{\a,\a^1,\a^2}(T)-\psi^{s,t}_{\a,\a^2,\a^1}(T)$.
We rewrite it as a polynomial map
$$
\Psi^{s,t} :  A\ot B\ot C \ra 
(\La 2(S^sA)\ot S^tA)\ot 
\La{s+t}B\ot \La{s+t}C
\ot \La sB\ot \La sC.
$$
So just as with Strassen's equations, we no longer need to make
choices of elements of $A^*$.

The only catch is we don't know whether or not  
$\Psi^{s,t}$ is identically
zero. In  \cite{LMsecb}
we show many of the $\Psi^{s,t}$ are indeed nonzero and give
independent subspaces
(in fact independent $GL(A)\times GL(B)\times GL(C)$-submodules,
see \S\ref{reptheorysect}) of the ideal of $\s_{s+t}(Seg(\BP A\times
\BP B\times \BP C))$.

In \cite{LMsecb}, Corollary 5.6,
using the above methods, we show that set-theoretic defining equations
for $\s_4(Seg(\pp 3\times \pp 3\times \pp 3))$,
 the
case of interest for phylogenetic invariants, 
could be explicitly determined if one had
a complete set of defining equations for
$\s_4(Seg(\pp 2\times \pp 2\times \pp 3))$.

\section{Representation theory and equations for secant varieties of Segre varieties}\label{reptheorysect}

As mentioned in the introduction, the most important tool
for studying varieties invariant under a group action is
representation theory. In this section we develop the necessary
represntation theory for studying secant varieties of 
Segre varieties. The theory developed in this section is
also what is needed in the more general study of algebraic
statistical models.
 We first describe how
to decompose the space of polynomials on
$A_1\otc A_n$ into subspaces invariant under the
action of the group of changes of bases in the vector spaces,
$GL(A_1)\ctimes GL(A_n)$. We then describe   Strassen's equations from this perspective
and how to
find preferred polynomials in each irreducible submodule.
We also describe two notions, {\it inheritance} and {\it prolongation},
which facilitate  our study.
Once one has an explicit description of a space of polynomials
as modules, it is algorithmic to write down an explicit
basis of the module as we did in \S\ref{warmup}. See
\cite{LMsec,LMor} for more details.

\subsection{Polynomials come in modules}
Since $\s_r(Seg(\BP A_1\times \cdots\times \BP A_n))$ is
invariant under the action of $G=GL(A_1)\times \cdots\times GL(A_n)$
acting on $ A_1\otc A_n = V$, its ideal, which is a 
subset of the module
$\oplus_d S^dV^*$, must be as well.
Thus we should study the equations of $\s_r(Seg(\BP A_1\times \cdots\times \BP A_n))$
as $G$-modules. 

Given any $G$-module $W$, the first thing to do when studying $W$
is to try to decompose it into {\it isotypic} components  (which
is always possible when $G$ is reductive, as is our situation). That is,
one can decompose $W$ into a direct sum of irreducible modules, but
this is not canonical. The isotypic decomposition
(which is canonical) is obtained
from the decomposition into irreducible submodules by grouping
together all copies of isomorphic irreducible submodules.

To decompose
$S^dV^*$ into $G$-isotypic components we   use 
the {\it Shur-Weyl duality}  between representations
of the symmetric group on $d$ letters $\FS_d$ and the representations
of the general linear group $GL(W)$. Both  groups act on
$W^{\ot d}$:
for $A\in GL(W)$ and $\s\in \FS_d$ we respectively have
\begin{align*}
A.(v_1\otc v_d)&=(A.v_1)\otc (A.v_d)\\
\s.(v_1\otc v_d)&= v_{\s(1)}\otc v_{\s(d)}
\end{align*}

Schur-Weyl duality is the
statement that each group is the commuting subgroup  of the other,
that is 
\begin{align*}
\FS_d&=\{g\in GL(W^{\ot d})\mid g.A.(v_1\otc v_d)
=A.g.(v_1\otc v_d) \ \forall A\in GL(W), \forall v_1\hd v_d\in W\}\\
&{\rm
and}\\
GL(W)&=\{g\in GL(W^{\ot d})\mid g.\s.(v_1\otc v_d)
=\s.g.(v_1\otc v_d) \ \forall \s\in \FS_d, \forall v_1\hd v_d\in W\}.
\end{align*}

Thus we can use the action of $\FS_d$ to obtain
projection operators $W^{\ot d}\ra W^{\ot d}$, whose
images are necessarily $GL(W)$-submodules. Moreover, the
duality assures us that all $GL(W)$-submodules
may be obtained this way. For example

\begin{align*}
S^dW&=\{ T\in W^{\ot d}\mid \s(T)=T \ \forall \s\in \FS_d\}\\
&=  \timage\pi_S: W^{\ot d}\ra W^{\ot d} \ {\rm where} \ \pi_S(w_1\otc w_d) 
=\frac 1{d!}\sum_{\s\in \FS_d} w_{\s (1)}\otc w_{\s (d)} \\
\La dW&=\{ T\in W^{\ot d}\mid \s(T)= sgn(\s)T \ \forall \s\in \FS_d\}
\\
&=  \timage\pi_{\Lambda}: W^{\ot d}\ra W^{\ot d} \ {\rm where}\  \pi_{\Lambda}(w_1\otc w_d) 
=\frac 1{d!}\sum_{\s\in \FS_d}sgn(\s) w_{\s (1)}\otc w_{\s (d)}\}
\end{align*}

Let  $\pi=(p_1\hd p_f)$ be a partition of $d$,
i.e., $p_1\geq\cdots\geq p_f$ and $p_1+\cdots +p_f=d$.
We use the notations $|\pi|=d$ and $l(\pi)=f$.

The irreducible representations of $\FS_d$ are indexed
by partitions of $d$; we let $[\pi]$ denote the module
induced by $\pi$. Here $[\pi]$ may be obtained by a choice of Young symmetrizer
$c_{\l}$ corresponding
to a choice of a   Young tableau associated 
to $\pi$  and applying the projection operator $c_{\l}$ to
the group algebra $\BC[\FS_d]$ (see, e.g., \cite{FH}, chapter four). 

Define  $S_{\pi}W:=\thom_{\FS_d}([\pi], W^{\ot d})$,
which is  an irreducible   $GL(W)$-module. 
The $GL(W)$-isotypic decomposition of $W^{\ot d}$ is
$W^{\ot d}=\oplus_{|\pi|=d}[\pi]\ot S_{\pi}W$.
The first factor is a trivial $GL(W)$-module so
it only serves to tell us the multiplicity of the second,
which is $\tdim [\pi]$.
 
We now return to the space we are interested in,
$V=A_1\otc A_n$ as a $G=GL(A_1)\ctimes GL(A_n)$-module:

\begin{proposition}[\cite{LMsec}] The  $G=GL(A_1)\ctimes GL(A_n)$
isotypic decomposition of $S^d(A_1\otc A_n)$ is 
$$S^d(A_1\otc A_n)= \bigoplus_{|\pi_1|=\cdots =|\pi_k|=d}([\pi_1]\otc [\pi_n])^{\FS_d} \ot
S_{\pi_1}A_1\otc S_{\pi_k}A_k,$$
where $([\pi_1]\otc [\pi_k])^{\FS_d}$ denotes the space of $\FS_d$-invariants
(i.e., instances of the trivial representation of $\FS_d$) in $[\pi_1]\ot \cdots\ot [\pi_n]$. \end{proposition}

The $([\pi_1]\otc [\pi_n])^{\FS_d}$ factor in the tensor
product just serves to tell us the multiplicity
of $S_{\pi_1}A_1\otc S_{\pi_k}A_k$, via its dimension.

\begin{proof}
We need to decompose $S^d(A_1\otc A_n)$ as
a $G=GL(A_1)\times\cdots\times GL(A_n)$-module. We have
$$
(A_1 \ot \cdots \ot A_n)^{\ot d}=
\bigoplus_{|\pi_j|=d}([\pi_1]\ot \cdots\ot [\pi_n])\ot ( S_{\pi_1}A_1\otc  S_{\pi_n}A_n)
$$
But $S^d(A_1\otc A_n)\subset (A_1\otc A_n)^{\ot d}$ is the set of elements invariant under the
action of $\FS_d$. (Here $\FS_d$ only acts on the $[\pi_j]$, it
leaves the $ S_{\pi_j}A_j$'s invariant.)
\end{proof}

Now we need a way to calculate $\tdim ([\pi_1]\otc [\pi_k])^{\FS_d}$.  This
can be done
  using {\it characters} in low degrees (degrees as high as your
computer is willing to tolerate).
The key point is
$$
\tdim ([\pi_1]\otc [\pi_n])^{\FS_d}  =\frac 1{d!}
\sum_{\s\in\FS_d} \chi_{\pi_1}(\s) 
  \cdots\chi_{\pi_n}(\s)
$$ 
where
$\chi_{\pi_j}: \FS_d\ra\BC$ is the {\it character} of $[\pi_j]$
(see, e.g., \cite{FH,Serre}).
For any given $d$, one can compute these dimensions,
but there is no known closed form formula for them
when $n>2$.

Obtaining the above decomposition   is essential
when dealing with explicit equations. 
For example, Strassen has {\it a priori}
three sets of equations for $\s_3(\pp 2\times \pp 2\times\pp 2)$.
Are they redundant or not? By examining these equations as modules
we find that they are:

\subsection{Strassen's equations as modules}\label{strassmodulesect}
Recall from \S\ref{invarstrassen}
that Strassen's equations for $\s_r(Seg(\pp 2\times\pp{b-1}\times\pp{b-1}))$ in degree $b+1$ are obtained
by composing the inclusion 
$$\La 2 A\ot S^{b-1}A\ot \La bB\ot B\ot C\ot \La bC\ra
(A\ot B\ot C)^{b+1}
$$ 
with the projection
 $$(A\ot B\ot C)^{b+1}\ra
S^{b+1}(A\ot B\ot C).
$$

Now $\La 2 A\ot S^{b-1}A\ot \La bB\ot B\ot C\ot \La bC$ is
not an irreducible module. Since the maps are $G$-equivariant,
by Shur's lemma the image is a direct sum of irreducible
submodules. We need to 
     determine
which modules in
$\La 2 A\ot S^{b-1}A\ot \La bB\ot B\ot C\ot \La bC$ map nontrivially
into $S^{b+1}(A\ot B\ot C)$.

Since here $b=\tdim B=\tdim C$, we have,
using a very special case of the Littlewood-Richardson rule
(see, e.g., \cite{FH}, chapter 6),
$$(\La 2 A\ot S^{b-1}A)\ot (\La bB\ot B)\ot (C\ot \La bC)
=(S_{b,1}A\op S_{b-1,1,1}A)\ot  \Lambda_{b,1}B \ot \Lambda_{b,1}C
$$
(where  we use the notation $\Lambda_{b,1}B=S_{2,1\hd 1}B$)
so there are two possible modules. Were the first in the image,
then one would be able to get equations in the case $\tdim A=2$,
but $\s_3(\pp 1\times \pp 2\times \pp 2)=\BP (A\ot B\ot C)$,
so only the second can occur (and it is easy to check that it does).
We conclude:

\begin{proposition}\cite{LMsecb} Strassen's 
equations for $\s_b(\pp 2\times \pp {b-1}\times \pp{b-1})$ 
expressed as a module is
$$S_{b-1,1,1}\BC^3\ot \Lambda_{b,1}\BC^b\ot \Lambda_{b,1}\BC^b,
$$
in particular it is an irreducible module.
\end{proposition}

When $b=3$, we obtain
$S_{211}A\ot S_{211}B\ot S_{211}C$
which occurs with multiplicity one in $S^4(A\ot B\ot C)$.
Thus,   despite the apparently different role
of $A$ from $B$ and $C$, in this case   - and {\it only} in this case -  exchanging
the role of $A$ with $B$ or $C$ yields the same space of
equations.  

\subsection{Highest weight vectors}
When we study modules of polynomials, it will be convenient
to have a \lq\lq best\rq\rq\ polynomial in the module. For example, since an
irreducible $G$-module in $S^dV^*$ is either entirely in or
out of the ideal of a $G$-variety $Z\subset \BP V$, it is sufficient to check
just a single polynomial in the module. In general, this
\lq\lq best polynomial\rq\rq\ 
is provided by a choice of highest weight vector.
We explain how to obtain such vectors when 
$G=GL(A_1)\ctimes GL(A_n)$.

Fix a  basis $e_1\hd e_n$ of a vector space $V$.
Let $W$ be an irreducible $GL(V)$-module occurring in $V^{\ot d}$
for some $d$. We say $w\in W$ is a {\it highest weight vector}
for $W$,  if $\rho (g).[w]=[w]$ for all upper triangular matrices
$g\in GL(V)$. (It makes sense
to discuss matrices because we have fixed a basis of $V$.)
  Highest weight vectors are in some sense
the simplest vectors occurring in a module.
For example, when  $W=S^dV$,   $(e_1)^d$ is a highest weight vector.
For $W=\La dV$, $\ee 1\ww\ee 2\ww\cdots\ww \ee d$ is a highest weight
vector. In general the highest weight vector of an irreducible
module will not correspond
to a decomposable tensor. In $c_{\pi}V^{\ot d}$ ($\simeq S_{\pi}V$), the
highest weight vector is 
$$c_{\pi}(e_1^{\ot p_1} \ot  e_2^{\ot p_2} \otc  e_d^{\ot p_d} )
$$
where $\pi=(p_1\hd p_d)$ and we allow the last few $p_j$ to be zero
in order to
have a uniform expression.

In \cite{LMsec} we give explicit
algorithms for writing down highest weight vectors
of submodules of  $S^d(A_1\otc A_n)$.

An important observation for the next section is if
$v\in A^{\ot d}$ is a highest weight vector for a submodule
corresponding to a partition $\pi$ and 
 $a_1\hd a_n$ is a basis of $A$, $v$ may be expressed
using only the vectors $a_1\hd a_{l(\pi)}$.

\subsection{Inheritance}\label{inheritsubsect}

By examining equations grouped into modules, the dimensions of the
vector spaces involved  only come into play when verifying that
the dimension is large enough to support a given module. For example:

\begin{proposition}\cite{LMsecb}\label{11.3} If a copy of
$$S_{\pi_1}A_1\ot \cdots S_{\pi_n}A_n$$
occurs in
$$
I_d(\s_r(Seg(\BP A_1^*\ctimes \BP A_n^*))),
$$
 then for all vector spaces $A_j'\supseteq A_j$,
the corresponding copy of 
$$S_{\pi_1}A_1'\ot \cdots\ot  S_{\pi_n}A_n'$$
occurs in 
$$
I_d(\s_r(Seg(\BP {A_1'}^*\times \cdots \times  \BP {A_n'}^*))).
$$

Moreover, a module $S_{\pi_1}A_1'\ot \cdots \ot S_{\pi_n}A_n'$
where the length of each $\pi_j$ is at most $\aaa_j$ is in
$I_d(\s_r(Seg(\BP {A_1'}^*\times \cdots \times  \BP {A_n'}^*)))$ if
and only if  the
corresponding module is in $I_d(\s_r(Seg(\BP A_1^* \times \cdots \times
\BP A_n^* ))$.
\end{proposition}

Our notation is such that given a variety
$Z\subset\BP V^*$, $I(Z)\subset S^\bullet V$ denotes
its ideal and $I_d(Z)=I(Z)\cap S^dV$.

\begin{proof}
A module is in the ideal if and only if  its highest weight vector is. Choose
ordered bases for $A_j'$ such that the first $\aaa_j$ basis vectors
form a basis of $A_j$. Then any highest weight vector for
$S_{\pi_1}A_1'\ot \cdots \ot S_{\pi_n}A_n'$ is also
a highest weight vector for
$S_{\pi_1}A_1 \ot \cdots \ot S_{\pi_n}A_n $
as long as $l(\pi_j)\leq \aaa_j$.
\end{proof}

{11.3} Thus a
copy of a module $S_{\pi_1}A_1\ot \cdots \ot S_{\pi_n}A_n$ will be
in $I(\s_r(Seg(\pp {r-1}\times \cdots \times \pp{r-1})))$
 if and only if  the corresponding copy of the module
$S_{\pi_1}\BC^{l(\pi_1)}\ot \cdots \ot S_{\pi_n}\BC^{l(\pi_n)}$ is
in the ideal of $\s_r(Seg(\pp {l(\pi_1)-1}\times \cdots \times
\pp{l(\pi_n)-1}))$.

It is straightforward to determine
$I_3(\s_2(Seg(\BP A_1\ctimes \BP A_n)))$ as a module:

\begin{theorem}[\cite{LMsec}, Theorem 4.7]\label{cubicsprop}
The space of cubics vanishing on  $\s_2(Seg(\BP A_1^*\times \cdots \times\BP A_k^*))$ is 
\begin{multline*}
 I_3(\s_2(Seg(\BP A_1^*\times \cdots \times\BP A_k^*)))  =  \bigoplus_{\substack{ I + J + L =
\{1\hd k\},\\ j=|J|>1,\, |L|>0}}
 \frac{2^{j-1}-(-1)^{j-1}}{3} S_3A_I\ot S_{21}A_J\ot S_{111}A_L \\
% & & \hspace{3cm} 
\oplus \bigoplus_{\substack{  I + J =\{1\hd k\},\\ j=|J|>3}}
(\frac{2^{j-1}-(-1)^{j-1}}3 -1)S_3A_I\ot S_{21}A_J
% & & \hspace{6cm} 
\oplus
\bigoplus_{\substack{ I + L =\{1\hd k\},\\ |L|>0\,even}} S_3A_I\ot S_{111}A_L.
\end{multline*}
\end{theorem}

\subsection{Prolongation}\label{prolongsect}
For $A\subset S^kV$ define $A\up p= (A\ot S^{p}V)\cap S^{p+k}V$,
the {\it $p$-th prolongation of $A$}. Let 
$$\tbase (A)=
\{ [v]\in \BP V^*\mid P(v)=0\ \forall P\in A\}.
$$
 
Ideals of secant varieties   satisfy a
{\it prolongation property}, in particular for secant varieties of intersections
of quadrics we have: 

\def\tbase{{\rm Zeros}}

\begin{lemma} \cite{LM0} \label{prolonglemm} Let $A\subset S^2V$ be a linear
subspace with zero set $\tbase (A)\subset \PP V^*$. 
Then $$\tbase (A\up{k-1})\supseteq \s_k(\tbase (A)).$$
Moreover, if $\tbase (A)$ is not contained in a hyperplane, then for $k\geq 2$, 
$I_k(\s_k(\tbase (A))=0$, and if $A=I_2(\tbase (A))$, then 
$I_{k+1}(\s_k(\tbase (A)))=A\up {k-1}$.
\end{lemma}

Usually,
for a variety $X\subset \BP V$,  $I(\s_k(X))$ is not generated in degree $k+1$. 
For example, consider the simplest intersection of quadrics, four points
in $\BP^2$. They generate six lines so $\s (X)$ is a hypersurface of
degree six.

\smallskip

Let $G$ be a semi-simple Lie or algebraic group, let 
$V_{\l}$ be the irreducible $G$-module of highest weight $\l$ 
 and let $X=G/P\subset \BP V_{\l}^*$ be a homogeneously
embedded rational homogeneous variety, i.e., 
the orbit of a highest weight
line.
($X=Seg(\BP A_1^*\otc \BP A_n^*)\subset \BP (A_1\otc A_n)^*=\BP V^*$
is one such.)
By an unpublished theorem of Kostant,   $I_2(X)=(V_{2\l}^*)\upperp \subset S^2V_{\l}$ and $I(X)$
is generated in degree two. More generally,     $I_k(X)=(V_{k\l}^*)\upperp
\subset S^kV_{\l}$. We adopt the notation that if
$V=V_{\l}$, we write $V^k=V_{k\l}$.
In the Segre case, 
$$
V^k =S^kA_1\otc S^kA_n\subset S^k(A_1\otc A_n)
$$

\begin{proposition}\cite{LMsec}\label{turbostuffin}
Let $X \subset \BP V^*$ be a  variety not contained
in a linear space. 
 Then for all $d>0$, $I_d(\s_d(X))=0$.
 
If $X=G/P$ is homogeneous, then  $I_{d+1}(\s_d(X))$ is the kernel of the contraction map $(V^2)^*\ot S^{d+1}V\ra S^{d-1}V$.
\end{proposition}

Examples illustrating Proposition \ref{turbostuffin} are given in
\cite{LMsec}. Extensions and further applications
of prolongations are given in \cite{sidman}.

\section{Auxiliary varieties}\label{auxvarsect}
A simple observation is that if $X\subset Y\subset \BP V$,
then any polynomial vanishing on $Y$ also vanishes
on $X$. We want to find polynomials
in the ideal of secant varieties of Segre varieties,
so it is natural to look for varieties
$Y$ that contain $X=\s_r(\BP A_1\ctimes \BP A_n)$ whose
ideals we   understand.
In this section we give two   examples
of such varieties $Y$.

\subsection{$Flat^{\overline a}_r$ and the GSS conjecture}
For example, note that $A\ot B\ot C=A\ot (B\ot C)$,
which leads to the simple observation that 
$\s_r(Seg(\BP A\times \BP B\times \BP C))\subseteq
\s_r(Seg(\BP A\times \BP (B\ot C)))$. Moreover
we explicitly know the generators of the
ideal of $\s_r(Seg(\BP A\times \BP (B\ot C)))$,
 see \S\ref{warmup}.

More generally, define the  {\it flattening}
of a tensor $T\in A_1\otc A_n$ by letting
to let $I=\{i_1\hd i_p\}\subset \{1\hd n\}$, $J=\{1\hd n\}\backslash I$,
$A_I=A_{i_1}\otc A_{i_p}$, $A_J=A_{j_1}\otc A_{j_{n-p}}$
and consider $T\in A_I\ot A_J$.

Let $\overline a=(a_1\hd a_n)$ and define 
$I_{Flat^{\overline a}_r}$ to be the ideal
generated by the modules $\La {r+1}A_I^*\ot \La {r+1}A_J^*\subset
S^{r+1}(A_1\otc A_n)^*$ as $I,J$ range over complementary
subsets of 
  $\{1\hd n\}$.
We let $Flat^{\overline a}_r$ denote the corresponding variety,
i.e., 
$$Flat^{\overline a}_r=\cap_{I,J}\s_r(Seg(\BP A_I\times \BP A_J)).
$$
We have
$\s_r(\BP A_1 \ctimes \BP A_n )\subseteq Flat^{\overline a}_r$.

The {\it GSS conjecture} \cite{GSS} is that
equality holds when $r=2$. Actually the conjecture is the stronger
statement that 
$I_{\s_2(\BP A_1 \ctimes \BP A_n )}
=I_{Flat^{\overline a}_2}$.
The weaker statement  that  equality holds
as sets
was proven in \cite{LMsec}.
It was also shown in \cite{LMsec} that the
conjecture holds when 
$\overline a=(a_1,a_2,a_3)$.
Since $\s_2(\BP A_1^*\ctimes \BP A_n^*) $ is reduced and irreducible,
and $Flat^{\overline a}_2 $ is irreducible, to
prove the conjecture  it would be sufficient to show $Flat^{\overline a}_2 $
is reduced. Using the methods outlined in \S\ref{weymansect},
it is possible to reduce the conjecture further to
showing that $Flat^{\overline a}_2$ is arithmetically Cohen-Macaulay,
see \cite{LWsecseg}.

In \cite{GSS}, a computer calculation is presented
that gives the dimensions of the minimal space of generators
of the ideals of $\s_2(Seg(\pp 1\times \pp 1\times \pp 1\times \pp 1))$
and $\s_2(Seg(\pp 1\times \pp 1\times \pp 1\times \pp 1\times \pp 1))$,
which, as shown in \cite{AR3},   allows one to
prove the GSS conjecture for up to five factors.
The proof relies on a variant of
\label{lmsecbprop} which was arrived at independently
using the geometry of phylogenetic trees.

\subsection{Subspace varieties}

\begin{definition}\label{subr}
Define the   {\it $s$-subspace variety}  
\begin{equation}
Sub_s:=\BP \{ T\in A\ot B\ot C \mid \exists A'\subset A, B'\subset B,
C'\subset C, \tdim A'=\tdim B'=\tdim C'=s, T\in A'\ot B'\ot C'
\}
\end{equation}
\end{definition}

Note that $\s_s(Seg(\BP A\times \BP B\times \BP C))\subseteq
Sub_s$, so the equations of $Sub_s$ are also equations for
$\s_s(Seg(\BP A\times \BP B\times \BP C))$.

\begin{proposition}\cite{LMsecb}\label{lmsecbprop} The ideal of $\s_r(Seg(\BP
A_1^*\times\cdots\times \BP A_n^*))$, when 
each $\tdim A_j^*\geq r$   is 
generated by the union of the the modules 
in its ideal inherited from 
the
modules generating the ideal of $\s_r(Seg(\pp {r-1}\times \cdots \times \pp{r-1}))$
and the modules generating the ideal of $Sub_{r}$. 
\end{proposition}

 To see this, note that by Proposition \ref{11.3}, a
copy of a module $S_{\pi_1}A_1\ot \cdots \ot S_{\pi_n}A_n$ will be
in $I(\s_r(Seg(\pp {r-1}\times \cdots \times \pp{r-1})))$
 if and only if  the corresponding copy of the module
$S_{\pi_1}\BC^{l(\pi_1)}\ot \cdots \ot S_{\pi_n}\BC^{l(\pi_n)}$ is
in the ideal of $\s_r(Seg(\pp {l(\pi_1)-1}\times \cdots \times
\pp{l(\pi_n)-1}))$.

The ideal of $Sub_r$ is easy to describe:

\begin{theorem}\cite{LWsecseg}\label{subrideal} The ideal of $Sub_{r}$ is generated
in degree $r+1$ by the   modules 
\begin{equation}\label{subrmodules}\La{r+1}A_j\ot \La{r+1}(A_1\otc A_{j-1}\ot A_{j+1}\otc A_n)
\end{equation}
for $1\leq j\leq n$
(minus   redundancies).
\end{theorem}

\begin{proof}First note that the ideal of
$Sub_{r}$ consists of all modules
$S_{\pi_1}A_1\otc S_{\pi_n}A_n$ occurring in $S^d(A_1\otc A_n)$
where each $\pi_j$ is a partition of $d$ and at least one $\pi_j$
has $l(\pi_j)>r$. We need to show that this ideal
is generated by the modules \eqref{subrmodules}.
But for each $j$,  the ideal consisting of representations 
$S_{\pi_1}A_1\otc S_{\pi_n}A_n$ occurring in $S^d(A_1\otc A_n)$
where  $l(\pi_j)>r$ is generated in degree $r +1$ by
$$\La{r+1}A_j\ot\La{r+1}(A_1\otc   A_{j-1}\ot A_{j+1} \otc A_n),
$$
because it is just the ideal of $\s_r(\BP A_j\times \BP(A_1\otc
\hat A_j\otc A_n))$.
\end{proof}

\begin{corollary}\cite{LMsec} The ideal of $\s_2(Seg(\BP A^*\times \BP B^*\times\BP C^*))$
is generated in degree three by
$\La 2A\ot \La 2(B\ot C)$,$\La 2B\ot \La 2(A\ot C)$ and 
$\La 2C\ot \La 2(A\ot B)$.
\end{corollary}
\begin{proof} $\s_2(\BP A\times \BP B\times \BP C)=Sub_2$
because $\s_2(\pp 1\times \pp 1\times \pp 1)=\BP (\BC^2\times\BC^2\times
\BC^2)$.
\end{proof}

We remark that the spaces $\La 2A\ot \La 2(B\ot C)$,$\La 2B\ot \La 2(A\ot C)$,
$\La 2C\ot \La 2(A\ot B)$ 
intersect, so there is redundancy in the above
description. This redundancy becomes apparent if one expresses the
spaces as sums of irreducible modules.

The $s$-subspace variety  is a cousin of the {\it rank varieties} in \cite{weyman}.    Moreover, it has a natural desingularization
explained in \S\ref{weymansect}.

\section{Weyman's method}\label{weymansect}

In this section we describe techniques for obtaining
generators of the ideals of secant varieties of
Segre varieties and more generally of $G$-varieties
$Z\subset \BP V$, where $G$ is a reductive group,
$V$ is a $G$-module and $Z$ is a variety
invariant under the action of $G$. In addition
to providing generators of the ideal, the techniques
enable one to compute the
entire minimal free resolution of the ideal of $Z$
as well as precise information about the
singularities of $Z$. These techniques
require  considerably more machinery
from commutative algebra and representation theory
than we have used up until this point. We expect they will 
be useful in future work.

Let $G$ be a reductive group, let $V$ be an irreducible
$G$ module, and let $Z\subset \BP V$ be a $G$-variety.

$G$-varieties are often uniruled by large linear spaces,
and   singularities occur when the linear spaces
crash into one another. To remedy this, one could
try to untangle the linear spaces. This appears
to be the idea underlying Kempf's
  desingularization by   the
{\it collapsing of a vector bundle}.
The idea is, given a $G$-variety
$Z\subset \BP V$, to find  (i.)  a homogeneous variety $G/P$, (ii.)  a homogeneous vector bundle  $E\ra G/P$ that
is the subbundle of a trivial bundle $\uV$ with fiber isomorphic to $V$
(here $P$ is a parabolic subgroup of $G$),
and (iii.) a map $\BP E\ra Z$ that is a desingularization.

For example, let $G(k,A)$ denote the Grassmannian
of $k$-planes through the origin in $A$. 
let $G=GL(A)\times GL(B)\times GL(C)$, let
$Z=Sub_s$ be as defined in \S\ref{subr}. Then
let $G/P=G(s,A)\times G(s,B)\times G(s,C)$ and let
$E=\cS_A\ot \cS_B\ot \cS_C$, where $\cS_A|_F$ is the 
$s$-plane $F\subset A$. Then $\BP E\ra Sub_s$ gives
the desired desingularization.

  Weyman takes Kempf's idea a step further  by observing that often
one can \lq\lq push down\rq\rq\  the minimal free resolution of the total space of
$E$ as a subvariety of the total space of the trivial bundle
(more precisely, of the structure sheaf of $E$ as an $\cO_{\uV}$-module)
  to obtain the minimal free resolution of 
$Z$.  Moreover, since the whole procedure is $G$-equivariant,
one gets the generators as modules.

The idea is as follows:
Assume  that the sheaf cohomology
groups $H^i(S^d(E^*))$ are all zero  for $i>0$ and for all $d$.
Consider the exact sequence
$$
0\ra (\uV/E)^*\ra \uV^*\ra E^*\ra 0
$$
giving rise, for each $j$, to a sequence
$$
0\ra \La j (\uV/E)^*\ra \La j\uV^* \ra \La {j-1}\uV^*\ot E^*\ra
\cdots \ra \uV^*\ot S^{j-1}E^*\ra S^jE^*\ra 0
$$

Since $\uV$ is trivial, and by our hypothesis 
all terms but the first have no cohomology in degree greater than
zero, 
  when we take the long exact sequence in cohomology, we
can split it into   short exact sequences that
we can in turn splice together to conclude
  that 
$
H^k(\La j(\uV/E)^*)
$
is the $k$-th homology of the sequence
$$
0\ra H^0(\La j \uV^*)\ra H^0(\La{j-1}\uV^*\ot E^*)\ra\cdots
\ra H^0(S^jE^*)\ra 0.  
$$
We add the hypothesis that the last step is surjective.

Now consider 
\begin{align*}
\La d\uV^* &\ra& \La{d-1}\uV^* \ot H^0(S^1E^*) &\ra &\cdots &\ra &
\uV^*\ot H^0(S^{d-1}E^*)&\ra&H^0(S^{d}E^*)&\ra&0\\
\uparrow & & \uparrow & & & & \uparrow & & \uparrow & &\\
\La d\uV^* &\ra& \La{d-1}\uV^* \ot V^* &\ra &\cdots &\ra &
\uV^*\ot S^{d-1}V^*&\ra&S^dV^*&\ra&0\\
\uparrow & & \uparrow & & & & \uparrow & & \uparrow & &\\
0 &\ra& \La{d-1}\ot  I_1(Z) &\ra &\cdots &\ra &
\uV^*\ot I_{d-1}(Z)&\ra&I_d(Z)&\ra&0
\end{align*}
where in the middle row we have $S^dV^* = H^0(S^d\uV^*)$ which
justifies the top row of vertical arrows. The horizontal
arrows are from the Koszul sequence.
The generators of the ideal of $Z$ in degree $d$ corresponds
to the cokernel of the lower right arrow. Now apply the
snake lemma to see it is the homology of the
$d$-th entry in the top sequence, which by the
observation above is $H^{d-1}(\La{d}(\uV/E)^*)$.
(One obtains the full minimal free resolution
in a similar fashion.)

All  the bundles in question are homogeneous. If they
are moreover irreducible, then one can apply the
Bott-Borel-Weil theorem to reduce the calculation of the
cohomology to a combinatorial calculation with the Weyl group
of $G$.
Even if they are not irreducible, one can use BBW on the
associated graded bundles and then apply spectral sequences. For
those who prefer to avoid spectral sequences in such
calculations, see \cite{OR}. 

Note that since we had to use the snake lemma, we have
no canonical way of identifying $H^{d-1}(\La{d}(\uV/E)^*)$
with the space of generators in degree $d$, but in the
equivariant setup, at least they agree as modules.

Sometimes it is sufficient to work with
a partial desingularization of $Z$, or a desingularization of a $G$ variety
that contains $Z$ as a variety of small codimension.

In fact, one does not need $Z$ to be a $G$-variety
(although for applications it almost always is).

\begin{theorem}\cite{weyman}\label{weymanthm}
 Let $Y\subset \BP V$ be a
variety and suppose there is a
projective variety $B$ and a vector bundle
$E\ra B$ that is a subbundle of a trivial bundle
$\underline V \ra B$ with $\underline V_z\simeq V$ for
$z\in B$ such that $  E\ra \hat Y$ is a desingularization.  Write $\eta=E^*$ and $\xi=(\underline V/E)^* $

If
the sheaf cohomology groups
$H^i(B,S^d\eta)$ are all zero for $i>0$
and   the linear maps
$H^0(B,S^d\eta)\ot V^*\ra H^0(B,S^{d+1}\eta)$
are surjective for all $d\geq 0$, then
\begin{enumerate}
\item
$\hat Y$ is normal, with rational singularities. 

\item The coordinate ring $K[\hat Y]$ satisfies
  $K[\hat Y]_d\simeq H^0(B,S^d\eta)$.

\item The 
vector space of minimal generators of the ideal of $\hat Y$ in
degree $d$ is isomorphic to  
$H^{d-1}(B,\La{d}\xi)$, which is
also the homology of the sequence
$$
\La 2 V\ot H^0(B,S^{d-2}\eta)\ra V\ot H^0(B,S^{d-1}\eta)
\ra H^0(B,S^d\eta).
$$

\item More generally,  
$\oplus_j H^j(\La{i+j}\xi)$ is isomorphic to
the $i$-th term in the minimal free resolution
of $Y$. 
\end{enumerate}

  If moreover $Y$ is a $G$-variety
and the desingularization is $G$-equivariant,
then the identifications above are as $G$-modules.

\end{theorem}

Using these methods, the minimal generators of the ideals
of $\s_r(Seg(\pp 1\times \pp b\times \pp c))$,
$\s_3(\pp a\times \pp b\times \pp c)$ and
$\s_2(\pp a\times \pp b\times \pp c\times \pp d)$ have been
  determined, see \cite{LWsecseg}. The method also gives
information about the singularities (e.g. normality, arithmetically
Cohen-Macaulay-ness), which, as mentioned above, can
be used to reduce problems such as the GSS conjecture.

\section{Appendix: Invariant formulations of two definitions from complexity
theory}\label{appendix}

The purpose of this section is to show how  {\it multiplicative
complexity} and {\it separations}  can be viewed invariantly,
and to discusses  advantages of the invariant perspective.
While the discussion is elementary, it is intended primarily
for those already familiar with these notions and their
uses.

\subsection{Multiplicative complexity and tensors}
A slightly larger class of algorithms for executing
bilinear maps $f:A\times B\ra C$ than those discussed in \S\ref{1.2}  is obtained
by writing $V=A\op B$ and considering $T$ as a bilinear
map $V\times V\ra C$. The {\it multiplicative complexity}
of $T$ is the rank of $T$ considered as a
bilinear  map $V\times V\ra C$.
See \S\ref{hopkerex} for an example of a tensor $T$
whose multiplicative complexity is less than $\trankc(T)$.

The multiplicative complexity is the minimal number multiplications
needed over all algorithms expressible as {\it straight line programs}, which is a class of algorithms that are intended
to model (classical) computer programs.
See \cite{BCS}, Definition 4.2 for a precise definition and
a proof of this statement.

 The multiplicative complexity of a map is
bounded both above  by its rank
  (obvious) and below by half the rank  (see \cite{BCS}, p354). So  
if one is only concerned with  the exponent of matrix
multiplication, one may restrict to the study of rank.

Our definition of multiplicative complexity gives
an immediate proof of (14.8) in \cite{BCS} which says that
$R(T)\geq {\rm multiplicative\  complexity}(T)\geq 2R(T)$.
To see this, note that
  $(A\op B)\ot (A\op B)\ot C
=A\ot B\ot C\op A\ot B\ot C \ot A\ot A\ot C \op B\ot B\ot C$;  so
any expression for $T$ in $(A\op B)^{\ot 2}\ot C$ of rank
$r$ projects to an expression for $T$ of rank at most
$2r$ in $A\ot B\ot C$ (and of course the projections
to $A\ot A\ot C$ and $B\ot B\ot C$ must be zero).

Here is an example where the multiplicative complexity
of a tensor is lower than its rank whose
presentation here also  illustrates
our definition.

\begin{example}\label{hopkerex}
Write $V=A\op B$.
The multiplicative complexity of $T\in A \ot B \ot C$ is
its rank considered as an element of $V\ot V\ot C$.
(This definition differs from those in the literature,
e.g., \cite{BCS} p. 352, but is equivalent.)
  Alekseyev 
\cite{MR780851}, building on work of Hopcroft and Kerr \cite{hk252},
showed that 
$\trank (M_{2,2,3})=11$, but Waksman \cite{MR0455534}  
give an explicit algorithm for $M_{2,2,3}$ that uses
 $10$ multiplications. Here is such an algorithm
   expressed as a tensor
in $(A\op B)\ot (A\op B)\ot C$:
\begin{align*}
M_{2,2,3}=
& \frac 12(a^1_1+b^2_1)\ot (a^1_2+b^1_1)\ot(   c^1_1  -   c^2_1 )
         + \frac 12(a^1_1+b^2_2)\ot(a^1_2+b^1_2)
 \ot(   c^1_2  +    c^2_1  +    c^2_3 )
\\
&+ \frac 12(a^1_1+b^2_3)\ot(a^1_2+b^1_3)\ot( c^1_3  -  c^2_3 ) + 
 (a^2_1+b^2_1)\ot(a^2_2+b^1_1)\ot c^2_1
\\
& 
         + \frac 12(a^2_1+b^2_2)\ot(a^2_2+b^1_2)\ot(-   c^2_1  +    c^2_2- 
   c^2_3)
 + (a^2_1+b^2_3)\ot(a^2_2+b^1_3)\ot c^2_3 
\\
&
+
 \frac 12(a^1_1-b^2_1)\ot(-a^1_2+b^1_1)\ot(   c^1_1 +   c^2_1 )
+\frac 12(a^1_1-b^2_2)\ot(-a^1_2+b^1_2)\ot(   c^1_2 -   c^2_1 -  c^2_3 )
\\
&
+\frac 12(a^1_1-b^2_3)\ot(-a^1_2+b^1_3)\ot(  c^1_3 +  c^2_3 )+
\frac 12(a^2_1-b^2_2)\ot(-a^2_2+b^1_2)\ot( c^2_1 +   c^2_2 +   c^2_3 ).
\end{align*}
\end{example}

\begin{remark} It might also be natural to consider expressions
of $T\in A\ot B\ot C$ in $(A\op B\op C)^{\ot 3}$, although it
is not clear how to encode such an object in a straight line
program. In any case, the savings would be at best by a factor
of $6$ by the same reasoning as in the paragraph above.
\end{remark}

\subsection{Separations of computations}\label{separationsect}

A standard technique   for showing lower bounds   (due to
Alder and Strassen \cite{AS}), 
is   {\it separations}.
The best known lower bound for $M_{3,3,3}$ is $19$ 
(due to Bl\"aser \cite{Bl2}). It  
is obtained by extensive use of separations. 
 In this section we define separations
in a more invariant fashion than  in \cite{AS}   and
suggest a more geometric   variant.

\begin{definition}Let $\phi\in A^*\ot B^*\ot C $ be
a computed tensor with computation of length $r$.
Let $A_1\subseteq A$, $B_1\subseteq B$, $C_1\subseteq C$
be subspaces. We say $\phi$ {\it separates} $(A_1,B_1,C_1)$
if we may write $\phi=\phi_1+\phi_2 +\phi_3$ where
the $\phi_i$'s are computed tensors whose lengths sum
to $r$ with the properties that
$$
\tlker (\phi_1|_{A_1})=0, \ \trker(\phi_2|_{B_1})=0 
$$
and no decomposable tensor appearing in 
the expression $\phi_1+\phi_2$ takes values
in $C_1$.
(This definition is equivalent to the standard one.)
Here for a bilinear map $\psi: A\times B\ra C$,
$\tlker (\psi)=\{ a\in A\mid \psi(a,b)=0\, \forall b\in B\}$
and similarly for $\trker(\psi)\subset B$.
\end{definition}

For $\phi$ as above, the length of $\phi$ is at least
$ \tdim A_1+\tdim B_1$ plus the number of decomposable tensors
appearing in $\phi_3$ taking values in $C_1$;
this is called the {\it Separation Lemma}.
As this observation indicates, separations are useful
for obtaining lower bounds for the rank of a tensor.

\smallskip

If $\tlker (\phi)=0$, then $\phi$ separates
$(A,0,0)$, and similarly for the right kernel. If
$\tim (\phi)=C$ then $\phi$ separates $(0,0,C)$.
Also, if $\phi$ separates $(A',B',C')$ then for
any $A''\subseteq A'$, $B''\subseteq B'$, $C''\subseteq C'$,
$\phi$ separates $(A'',B'',C'')$.

\smallskip

\begin{lemma}[Extension lemma]\cite{AS}
Let $\phi\in A^*\ot B^*\ot C$ be a computed tensor  that separates $(A_1,B_1,C_1)$.
 Let $A_1\subseteq A_2\subseteq A$.   If
$\phi$ fails to separate  $(A_2,B_1,C_1)$, then there exists
$a\in A_2\backslash A_1$ with 
\begin{equation}\label{sepeqn}\phi(a,B)\subseteq 
  \phi (a,B_1)  + C_1.
\end{equation}
\end{lemma}

Of course the same is true with the roles of $A$ and $B$ interchanged.

\begin{proof} 
We try to write $\phi=\tilde\phi_1+\tilde\phi_2+\tilde\phi_3$ such that the tilded splitting of $\phi$ separates $(A_2,B_1,C_1)$.

Write $\phi_3=\tilde \phi_3+\phi_3'$ with
$\tim (\tilde\phi_3)\subset C_1$ and $\tilde\phi_3$
maximal with this property. (Note that $\tilde \phi_3$ is unique.)
Then consider $\psi =\phi_1+\phi_2+\phi_3'$ and say $\psi$
has length $l$. Then we have the best chance of separating
$(A_2,B_1,C_1)$ if we choose $\tilde \phi_2$
of  minimal rank such that
$\trker \tilde\phi_2\mid_{B_1}=0$. Thus the length
of $\tilde \phi_2=\tdim B_1=:b_1$. There are at most
$\binom l{b_1}$ choices of such $\tilde\phi_2$. Given
any admissible such choice, the resulting $\tilde \phi_1:=
\psi-\tilde\phi_2$ must also have the property
that $\tlker\tilde\phi_1\mid_{A_1}=0$. Say we have such
a choice and we want to see if the separation extends
to $A_2$, i.e., that $\tlker\tilde\phi_1\mid_{A_2}=0$.
Now suppose not, then there exists $a\in A_2\backslash A_1$
such that $a\in \tlker (\tilde\phi_1)$, and thus
for all $b\in B$
$$
\phi(a,b)=\tilde\phi_2(a,b)+\tilde\phi_3(a,b).
$$
Write $B=B_1\oplus\trker(\tilde\phi_2)$ and
given $b\in B$, $b=b'+b''$ uniquely with $b'\in B_1$,
$b''\in \trker(\tilde\phi_2)$. So
$$
\phi(a,b)= \phi(a,b') +\tilde\phi_3(a,b'')
\in \langle \phi(a,B_1)\rangle +C_1
$$
So we see if $\phi$ fails to separate for at least one choice of
tilded splitting equation, then \eqref{sepeqn} holds. In particular equation \eqref{sepeqn} holds if it fails
for all possible choices.
\end{proof}

\smallskip

Here is an easy application of the extension lemma:

\begin{proposition}\label{ra0sep} If $A$ is a simple algebra and $R\subset A$
a maximal right ideal, then
any computation of $Mult_A$ separates $(R,A,0)$.
\end{proposition}

\begin{proof}
Since $\phi$ separates $(A,0,0)$ it separates $(R,0,0)$.
Let $B_1\subset B$ be maximal such that 
$\phi$ separates
$(R,B_1,0)$. If $B_1\neq B$ then there exists a nonzero
$b\in B$ such that $Ab\subseteq \langle RB\rangle =R$,
a contradiction as a left ideal cannot be contained in a right
ideal.
\end{proof}

As a corollary we obtain a very easy proof that
$\trankc (M_{m,m,m})\geq 2m^2-m$.

\begin{definition}
A more natural and general definition of separation (which, to 
avoid confusion, we call Separation), is as follows:
Given $T\in V_1^*\ot \cdots \ot V_n^*$,
$\phi$ a computation of $T$ and $U_j\subseteq V_j$
we will say $\phi$ {\it Separates} $(U_1\hd U_n)$ if we have a
decomposition
$\phi=\phi_1+\cdots + \phi_n+\psi$ with each
$$ \phi_j: U_j
\ra V_1^*\ot \cdots \ot V_{j-1}^*\ot V_{j+1}^*\ot \cdots
\ot V_n^*
$$
 injective and $\tlength (\phi)=\sum_i\tlength (\phi_i)+\tlength (\psi)$.
\end{definition} 

If $\phi$ Separates $(A_1,B_1,C_1)$ then the length of
$\phi$ is at least $ \tdim A_1+\tdim B_1+\tdim C_1$ so the
conclusion of the corresponding 
Separation lemma is a little stronger than
that of  the separation lemma
(but the hypotheses are stronger as well). Note that the
hypotheses are also basis independent, unlike the separation lemma.
 
We leave the statement and proof of the analogous
Extension lemma to the reader.

\bibliographystyle{amsplain}
%\nocite{*}
\bibliography{Lmatrix}

\end{document}

%% file: cortdefs.tex
\newtheoremstyle{custom}% name
  {3pt}%      Space above
  {3pt}%      Space below
  {\slshape}%         Body font
  {}%         Indent amount (empty = no indent, \parindent = para indent)
  {\bfseries}% Thm head font
  {.}%        Punctuation after thm head
  { }%     Space after thm head: " " = normal interword space;
   {}%         Thm head spec (can be left empty, meaning `normal')
\theoremstyle{custom}
\newtheorem{theorem}{Theorem}[section]

\newtheorem{proposition}[theorem]{Proposition}
\newtheorem{proposition/definition}[theorem]{Proposition/Definition}
\newtheorem{lemma}[theorem]{Lemma}
\newtheorem{corollary}[theorem]{Corollary}

\theoremstyle{definition}
\newtheorem{definition}[theorem]{Definition}

\newtheorem{example}[theorem]{Example}

\theoremstyle{remark}
\newtheorem{remark}[theorem]{Remark}

\def\donote#1{\noindent{\bf #1\ }}% for when nothing else works

\newtheoremstyle{exercise}% name
  {3pt}%      Space above
  {6pt}%      Space below
  {}%         Body font
  {}%         Indent amount (empty = no indent, \parindent = para indent)
  {\bfseries}% Thm head font
  {:}%        Punctuation after thm head
  { }%     Space after thm head: " " = normal interword space;
   {}%         Thm head spec (can be left empty, meaning `normal')
\theoremstyle{exercise}
\newtheorem{exercise}[theorem]{Exercise}
\newtheoremstyle{exercises}% name
  {3pt}%      Space above
  {6pt}%      Space below
  {}%         Body font
  {}%         Indent amount (empty = no indent, \parindent = para indent)
  {\bfseries}% Thm head font
  {:}%        Punctuation after thm head
  {\newline}%     Space after thm head: " " = normal interword space;
   {}%         Thm head spec (can be left empty, meaning `normal')

\theoremstyle{exercise}
\newtheorem{exercises}[theorem]{Exercises}

\input epsf
\def\boxit#1{\vbox{\hrule height1pt\hbox{\vrule width1pt\kern3pt
  \vbox{\kern3pt#1\kern3pt}\kern3pt\vrule width1pt}\hrule height1pt}}

\def\trank{\text{rank}}
\def\ee#1{e_{#1}}

\def\BC{\mathbb C}\def\BO{\mathbb O}
\def\BR{\mathbb R}
\def\BP{\mathbb P}
\def\pp#1{\mathbb P^{#1}}

\def\pp#1{{\mathbb P}^{#1}}
\def\tdim{{\rm dim}}

\def\hd{,...,}
\def\ww{\wedge}
\def\upperp{{}^\perp}

\def\na{n+a}

\def\inv{{}^{-1}}

\def\cS{{\mathcal S}}

\def\cO{{\mathcal O}}

\def\11{\mathbf 1}
\def\PP{\mathbb P}

\def\l{\lambda}
\def\a{\alpha}

\def\o{\omega}

\def\b{\beta}

\def\s{\sigma}

\def\d{\delta}

\def\up#1{{}^{({#1})}}

\def\ot{{\mathord{ \otimes } }}
\def\op{{\mathord{\,\oplus }\,}}
\def\otc{{\mathord{\otimes\cdots\otimes}\;}}

\def\ra{{\mathord{\;\rightarrow\;}}}

\def\La#1{\Lambda^{#1}}

\def\op{\oplus}
\def\BO{\Bbb O}

\def\bb#1#2#3{b^{#1}_{{#2}{#3}}}

\def\ep{\epsilon}
\def\op{\oplus}

\def\s{\sigma}
\def\t{\tau}

\def\a{\alpha}
\def\b{\beta}

\def\l{\lambda}

\def\FS{\mathfrak  S}

\def\BP{\mathbb  P}
\def\BC{\mathbb  C}

\def\pp#1{\mathbb  P^{#1}}

\def\tcodim{\text{codim}}
\def\BR{\mathbb  R}

\def\ep{\epsilon}

\def\aaa#1#2{a^{#1}_{#2}}

\def\cc#1#2{C^{#1}_{#2}}
\def\ccc#1#2#3{c^{#1}_{{#2}{#3}}}

\def\ee#1{e_{#1}}

\def\hd{, \hdots ,}

\def\inv{{}^{-1}}

\def\La#1{\Lambda^{#1}}

\def\pp#1{\mathbb  P^{#1}}

\def\ra{\rightarrow}

\def\tbase{\operatorname{Baseloc}}
\def\tdeg{\operatorname{deg}}
\def\tdet{\operatorname{det}}

\def\tim{\operatorname{Image}}
\def\tdim{\operatorname{dim}}

\def\tlim{\lim}

\def\tmin{\operatorname{min}}

\def\thom{\operatorname{Hom}}
\def\trank{\operatorname{rank}}

\def\up#1{{}^{({#1})}}
\def\upperp{{}^{\perp}}

\def\ww{\wedge}

\def\na{{n+a}}

\def\ctimes{\times \cdots\times}
\def\wcdots{\ww \cdots\ww}

\def\timage{\operatorname{Im}}